\documentclass[prc,preprint,showpacs,showkeys,superscriptaddress,nofootinbib,tightenlines]{revtex4}

\usepackage{graphicx,color}  %
\usepackage{bm}  %

\newcommand{\beq}{\begin{equation}}
\newcommand{\eeq}{\end{equation}}
\newcommand{\bea}{\begin{eqnarray}}
\newcommand{\eea}{\end{eqnarray}}
\newcommand{\ben}{\begin{eqnarray*}}
\newcommand{\een}{\end{eqnarray*}}

\newcommand{\simge}{\hspace*{0.2em}\raisebox{0.5ex}{$>$}
     \hspace{-0.8em}\raisebox{-0.3em}{$\sim$}\hspace*{0.2em}}
\newcommand{\simle}{\hspace*{0.2em}\raisebox{0.5ex}{$<$}
     \hspace{-0.8em}\raisebox{-0.3em}{$\sim$}\hspace*{0.2em}}

\begin{document}


\title{An effective field theory approach to two trapped particles}

\author{I. Stetcu}
\affiliation{Department of Physics, University of Washington, Box 351560, 
Seattle, WA 98195-1560}

\author{J. Rotureau}
\affiliation{Department of Physics, University of Arizona, 
Tucson, AZ 85721}

\author{B.R. Barrett}
\affiliation{Department of Physics, University of Arizona, 
Tucson, AZ 85721}

\author{U. van Kolck}
\affiliation{Department of Physics, University of Arizona, 
Tucson, AZ 85721}


\begin{abstract}
We discuss the problem of two particles interacting via short-range 
interactions within a harmonic-oscillator trap.
The interactions are organized according to their number of derivatives
and defined in truncated model spaces made from a bound-state basis.
Leading-order (LO) interactions are iterated to all orders, 
while corrections 
are treated in perturbation theory. 
We show explicitly that next-to-LO and next-to-next-to-LO interactions improve
convergence as the model space increases.
In the large-model-space limit we regain results from
a pseudopotential.
Arbitrary scattering lengths are considered, as well as 
a generalization to include the non-vanishing range of the interaction. 
\end{abstract}

\smallskip
\maketitle  

\section{Introduction}
\label{sec:intro}
Much experimental progress has recently been achieved in the area of 
ultra-cold atoms. It is now possible  \cite{kohl} 
to make systems of atoms 
confined in optical lattices formed by laser beams and at the same
time control the strength of the two-body interaction with magnetic fields.
The atoms are cooled down to extremely low temperatures
and, in the limit of low tunneling, each lattice site 
may be regarded as a harmonic-oscillator (HO) well,
which is independent from the others
and contains only a few atoms.
Using a Feshbach resonance,
the two-body interaction can be fine-tuned so that
the $S$-wave scattering length $a_2$ is made much larger than the
range $R$ of the interaction.
By doing so, the theoretical problem of a system of trapped particles 
interacting via short-range forces can be realized experimentally.

Such atomic systems present remarkable similarities to some nuclear physics 
systems.
Indeed, the two-nucleon scattering length is much larger than the range 
of the nuclear force, set by the pion mass. 
In such a situation, the physical properties of few-body systems 
are universal, that is, they depend 
mainly on $a_2$,
not on the details of the two-body interaction. 
At sufficiently low energies, an effective field theory (EFT) 
has been formulated 
which turns this separation of scales into an expansion 
in powers of $R/a_2$ \cite{aleph}.
Except for isospin, which does play a role in the relative relevance
of few-body forces, 
the version of this EFT used in nuclear physics \cite{nukeEFTrev}
is formally indistinguishable from 
the theory describing atomic systems \cite{Braat}. 
(Nevertheless, the underlying theories for the two cases are very different.)
As a consequence, atomic systems characterized by large scattering
lengths can be studied with techniques developed in nuclear physics
and, conversely, provide an excellent testing ground for few- and many-body 
methods that can be further applied, with little or no change, to the 
description of nuclear systems at low energies. 

The no-core shell model (NCSM) is one of the most flexible 
\textit{ab initio} methods used 
to obtain the solution to the non-relativistic Schr\"odinger equation for 
many-nucleon systems \cite{tradNCSM}. 
Currently, it is the only such 
method able to 
reach medium-mass nuclei with no restrictions to closed or nearly-closed 
shells and at 
the same time to handle local and non-local interactions on the same footing.  
It uses a discrete single-particle basis ---typically a HO basis--- and,
being a numerical method, relies on a suitable truncation of the 
model space accessible to nucleons ---in the form of a
maximum number $N_{max}$ of accessible shells above
the minimum configuration. 
This requires the use of effective interactions 
(not only among nucleons but also among nucleons and external probes such 
as photons, when they are present)
to account for effects left out by the truncation to a finite model space. 
The method of choice has been 
the construction of effective operators via unitary transformations. 
This involves the 
use of an approximation, the so-called ``cluster approximation", which is not 
\textit{a priori} 
controlled, and which poses challenges for the description of low-momentum 
observables \cite{ncsm:eff_op}.
An alternative is to use EFT to
construct effective interactions that are consistent with
the underlying theory of QCD directly in the model spaces where
many-nucleon calculations are carried out \cite{NCSM}.

In this paper, we consider the problem of two trapped atoms 
with large scattering length, $|a_2|\gg R$,
from the perspective of EFT interactions
solved with the NCSM.
The main simplification with respect to the nuclear case
is that the trapping potential provides a natural single-particle basis;
its HO length $b$ does not need to be removed at
the end of the calculation because it represents 
the long-distance physics of the trap.
As long as $b \gg R$, the trapped system should still exhibit universal 
behavior, although for $b \simle |a_2|$ it is significantly different from 
that of the untrapped system. 

We use EFT techniques to expand the interaction between particles as 
a series of contact interactions with an increasing number of derivatives,
and calculate energy levels inside the trap by explicitly solving the 
Schr\"odinger equation.
We extend 
here the work initiated in Ref. \cite{trap0}, in which only 
the leading-order (LO) interaction was investigated. 
We include explicitly corrections 
up to the next-to-next-to-leading-order (N$^2$LO),
and show how they provide improved convergence
as $N_{max}$ increases,
at least as long as they are treated in perturbation theory.
Finite and infinite values of the scattering 
length $a_2$ are considered,
and 
we show that in the $N_{max}\to \infty$ limit our LO results converge
to the levels of a pseudopotential \cite{pseudo}
with the same scattering length,
obtained by Busch {\it et al.} \cite{HT}.
Starting at NLO 
a finite value of the effective range $r_2$ is allowed as well,
and at higher orders other effective-range expansion parameters 
could be included similarly. 
The $N_{max}\to \infty$ limit of our subleading-order EFT 
provides a derivation of the generalized
Busch {\it et al.} relation \cite{models,DFT_short,mehen}.
With the two-body system thus understood, we can use our method
to calculate the energies of larger trapped systems \cite{usnext}.

Other approaches to the same problem of course exist in the literature.
They are frequently based on a specific form for the 
interparticle potential (see, for example, Ref. \cite{models}).
Closest in spirit to ours is probably the approach \cite{alhassid} where 
an effective short-range interaction is fitted to several levels
of the pseudopotential at unitarity, but diagonalized exactly.
What distinguish our framework are its
{\it i)} systematic character, in the form of a controlled expansion;
and
{\it ii)} generality,
since no assumptions about the form of the short-range interactions 
are needed.
These features allow the same method to apply across fields.
Of course, the low-energy observables themselves should be in agreement 
among different approaches, as long as they are calculated properly.

The paper is organized as follows. 
We first show in Sec. \ref{sect2} the general formalism of our approach, 
including a 
detailed description of the treatment beyond leading order. 
We illustrate the 
method in different situations (finite and infinite scattering length, 
negligible and 
non-vanishing effective range, {\it etc.}) 
in Sec. \ref{sect3}.
We conclude 
and discuss 
future applications in Sec. \ref{sect4}.
Appendices \ref{AppD}, \ref{appA}, and \ref{wf_NLO} 
provide some of the details omitted in the main text.

\section{General Considerations} 
\label{sect2}

We consider a non-relativistic system of two 
particles of reduced mass $\mu$ that interact with each
other in the $S$ wave.
For definiteness, we think of two-component fermions,
which support a single $S$ channel and interact also in $P$ and higher waves,
but the framework can be 
straightforwardly applied to bosons and other fermions.
In free space, the properties of this system
can be characterized by scattering phase shifts 
for each partial wave of angular momentum $l$ at relative 
on-shell momentum $k$, 
$\delta_l(k)$. 
When the relative momentum is much smaller than the inverse of the
range $R$ of the interparticle interaction, $k\ll 1/R$ 
(we use units in which $\hbar=1$ and $c=1$), the phase shifts 
are given
by the effective range expansion (ERE);
for example, in the $S$ wave,
\begin{equation}
k \, \cot \delta_0(k) =-\frac{1}{a_2}+\frac{1}{2}r_2 k^{2}
+\frac{1}{4}P_2 k^{4}+ \ldots,
\label{ERE}
\end{equation}
where 
$a_2$, $r_2$, $P_2$, $\ldots$ are, respectively,
the scattering length, effective range,
shape parameter, and higher ERE parameters not shown explicitly.
A similar expansion exists for $P$ and higher partial waves.

Generically, the sizes of ERE parameters are set by $R$, 
for example $|r_2|\sim R$. 
The ERE (\ref{ERE}) is an expansion in powers of $kR$.
Given a desired precision, the ERE can be truncated and
the system described by a finite number of parameters.
Potentials that generate the same values for this finite number
of ERE parameters cannot be distinguished at this precision level:
they all generate the same wavefunction for distances beyond
the range of the force,  $r\simge R$.
Such potentials are said to belong to the same universality class.
Most interesting is the case
where the depth of the interaction potential is fine-tuned
so that an $S$-wave bound state is near threshold
and the associated scattering length is large, $|a_2|\gg R$. 
Then the physics of the bound state is largely independent 
of the potential; 
for example, for $R\ll r\ll a_2$ the wavefunction of a state
of energy $E$ is
\begin{equation}
\psi(\vec{r})\propto 
\frac{1}{r}\left\{1-\left[1-\mu a_2 r_2 E+\ldots\right]\frac{r}{a_2}
+{\cal O}\left(\frac{r^2}{a_2^2}\right)\right\}.
\label{aswf}
\end{equation} 
Physics in the unitarity limit, $|a_2|\to \infty$, 
is controlled by the zero-energy bound state and exhibits
a higher degree of universality. 

The ERE description is useful because it is model independent.
However, it does not directly provide a basis for the study of many-body
systems. This can be accomplished using the ideas of EFT 
\cite{nukeEFTrev,Braat}.
Since at low momenta the  details of the interaction
cannot be resolved, 
we can expand the interparticle 
interaction $V$ 
as a Taylor series in momentum space,
as done in App. \ref{AppD};
in coordinate space,
\begin{eqnarray}
V(\vec{r}\,', \vec{r})
&=&C_0 \delta(\vec{r}\,') \delta(\vec{r})
  -C_2\left\{\left[\nabla\,'^2\delta(\vec{r}\,')\right] \delta(\vec{r})
    +\delta(\vec{r}\,') \left[\nabla^2\delta(\vec{r})\right]\right\}
         \nonumber\\
&&+ C_4 \left\{\left[\nabla\,'^4\delta(\vec{r}\,')\right] \delta(\vec{r})
	     +\delta(\vec{r}\,') \left[\nabla^4\delta(\vec{r})\right]
  + 2 \left[\nabla\,'^2\delta(\vec{r}\,')\right]
        \left[\nabla^2\delta(\vec{r})\right] \right\}   
+\ldots \label{Taylorcoord}
\end{eqnarray}
where $C_0$, $C_2$, 
and $C_4$ are parameters,
and ``\ldots'' denote interactions that contribute
at higher orders.
Because the interactions are singular, an 
ultraviolet (UV) cutoff $\Lambda$
has to be introduced to solve the Schr\"odinger equation.
In order for observables to be independent of $\Lambda$
(renormalization-group invariance),
the parameters $C_i$ have to depend on $\Lambda$.
This is merely a consequence of the fact that the UV cutoff
is an arbitrary separation between the short-range dynamics
included explicitly in the dynamics (through high virtual momenta)
versus that included implicitly in the potential (through its parameters).

Contributions to observables obtained from $V$ (\ref{Taylorcoord})
can also be organized
in powers of $kR$.
It has, in fact, been shown \cite{aleph} that in the two-body
sector the EFT expansion reproduces
the ERE (\ref{ERE}) at each power of $kR$.
In the generic situation, one can simply treat the whole potential
in perturbation theory. 
When $|a_2|\gg R$, however,
the $C_0$ term in Eq. (\ref{Taylorcoord}) needs to be solved
exactly, while the remaining terms can still be accounted for in 
perturbation theory
\cite{aleph}.
These higher-order terms represent range and subtler effects
in the $S$ wave, and $P$ and higher waves.

Here we are interested in the large-scattering-length scenario,
when the two particles are
trapped in an HO potential of frequency $\omega$.
The HO introduces a third scale,
the length $b=1/\sqrt{\mu\omega}$.
In the relative frame, the  Hamiltonian for this system reads
\begin{equation}
H= \frac{\omega}{2}\left[-b^2\nabla^2 + \frac{r^2}{b^2}\right] + V.
\label{hamil}
\end{equation}
As long as $b\gg R$, details of the interparticle potential remain
irrelevant and universality is not destroyed.
In the following we consider various values for the ratio $b/a_2$,
which are, as discussed in Sec. \ref{sec:intro}, 
of interest in both atomic and nuclear physics.

We want to set up a perturbative approach;
we thus write the Hamiltonian (\ref{hamil}) for 
the relative motion as
\begin{equation}
H=H^{(0)}+V^{(1)}+V^{(2)}+\ldots,
\end{equation}
with the energy and wavefunction decomposed accordingly,
\begin{equation}
|\psi\rangle =|\psi^{(0)}\rangle +|\psi^{(1)}\rangle +|\psi^{(2)}\rangle 
+\ldots,
\label{wf}
\end{equation}
and 
\begin{equation}
E=E^{(0)}+E^{(1)}+E^{(2)}+\ldots
\label{E}
\end{equation}
The superscript $^{(n)}$ corresponds to the order in perturbation theory of 
the different terms.
Here we order interactions according to the power counting of Ref. \cite{aleph}
and
for convenience we split the parameters $C_i$ in Eq. (\ref{Taylorcoord}) 
among different orders. 

The leading-order (LO) Hamiltonian is 
\begin{equation}
H^{(0)}=\frac{\omega}{2}\left[-b^2\nabla^2 + \frac{r^2}{b^2}\right]
        + C_0^{(0)} \delta(\vec{r}),
\label{LOpot}
\end{equation}
and the corresponding wavefunction $\psi^{(0)}(\vec{r})$  
is the solution of the Schr\"odinger equation
\begin{eqnarray}
\left(H^{(0)}-E^{(0)}\right)\psi^{(0)}(\vec{r})&=&0.
\label{sch0}
\end{eqnarray}
The next-to-leading-order (NLO) correction to the potential is
\begin{equation}
V^{(1)}=C_0^{(1)} \delta(\vec{r})
  -C_2^{(1)} \left\{ \left[\nabla^2\delta(\vec{r})\right]
             +2\left[\vec{\nabla} \delta(\vec{r})\right]\cdot \vec{\nabla}
             +2\delta(\vec{r})\nabla^2\right\},
\label{NLOpot}
\end{equation}
and the first-order corrections to the energy, $E^{(1)}$,
and to the wavefunction, $\psi^{(1)}(\vec{r})$,
are obtained in first-order perturbation theory. That is,
they are such that
\begin{eqnarray}
\left(H^{(0)}-E^{(0)}\right) \psi^{(1)}(\vec{r})
&=&\left(E^{(1)}-V^{(1)}\right)\psi^{(0)}(\vec{r}).
\label{sch1}
\end{eqnarray}
The next-to-next-to-leading-order (N$^2$LO) correction to the potential 
$V^{(2)}$ is given by
\begin{eqnarray}
V^{(2)}&=&C_0^{(2)} \delta(\vec{r})
  -C_2^{(2)} \left\{ \left[\nabla^2\delta(\vec{r})\right]
             +2\left[\vec{\nabla} \delta(\vec{r})\right]\cdot \vec{\nabla}
             +2\delta(\vec{r})\nabla^2\right\}
\nonumber\\
&&
+C_4^{(2)} \left\{\left[\nabla^4\delta(\vec{r})\right] 
         +4\left[\vec{\nabla} \nabla^2\delta(\vec{r})\right]\cdot \vec{\nabla}
         +4\left[\vec{\nabla} \vec{\nabla} \delta(\vec{r})\right]
                 \cdot \cdot \vec{\nabla}\vec{\nabla}
\right.\nonumber\\
&&\left. \qquad\quad
         +4\left[\vec{\nabla} \delta(\vec{r})\right]\cdot \vec{\nabla}\nabla^2
+2\delta(\vec{r})\nabla^4\right\}.
\label{NNLOpot}
\end{eqnarray}
The corrections $E^{(2)}$ and $\psi^{(2)}(\vec{r})$ are obtained 
from
\begin{eqnarray}
\left(H^{(0)}-E^{(0)}\right) \psi^{(2)}(\vec{r})
&=&\left(E^{(2)}-V^{(2)}\right)\psi^{(0)}(\vec{r})
   +\left(E^{(1)}-V^{(1)}\right)\psi^{(1)}(\vec{r}),
\label{sch2}
\end{eqnarray}
which just means perturbation theory to first order in $V^{(2)}$
and to second order in $V^{(1)}$.
Extension to higher orders is straightforward.

We work on a basis of HO wavefunctions $\phi_{nlm}(\vec r)$ 
with energies 
$E_{nl}=(2n+l+3/2)  \omega\equiv (N+3/2) \omega$.
Useful properties of these wavefunctions are 
summarized in App. \ref{appA}.
In the HO basis the singularity of the potential (\ref{Taylorcoord}) 
can be tamed by imposing a maximum number of
shells $N_{max}$ \cite{NCSM}, which corresponds to a UV momentum cutoff
\begin{equation}
\Lambda=\frac{1}{b}\sqrt{2N_{max}+3}.
\label{Lambda}
\end{equation} 
We therefore expand the wavefunction (\ref{wf})
in the HO basis in the finite space,
\begin{equation}
\psi^{(\nu)}(\vec{r})=
\sum_{n,l=0}^{N_{max}}\sum_{m=-l}^{l} 
c_{nlm}^{(\nu)} 
\; \phi_{nlm}(\vec{r}),
\label{expansion}
\end{equation}
where $c_{nlm}$ are coefficients to be determined. 

Since the two-body potentials up to N$^2$LO only support $S$ waves, 
to this order the eigenfunctions 
with $l>0$ are simply the HO wavefunctions $\phi_{nlm}(\vec r)$
with eigenvalues $(2n+l+3/2) \omega$ up to the energy 
$(N_{max} +3/2)\omega$.
For $l=0$, on the other hand, the levels are affected by
the interparticle potential. Denoting by $n_{max}$ the maximum
value of the radial quantum number (that is, $n_{max}$ is the largest
integer smaller than, or equal to, $N_{max}/2$)
and omitting the labels $l=0$ and $m=0$, the $S$ wavefunction
can be written as
\begin{equation}
\psi^{(\nu)}_0(r) =\sum_{n=0}^{n_{max}} c_n^{(\nu)} \phi_{n}(r)
\label{expansionS}
\end{equation}
in terms of the HO $S$ wavefunctions $\phi_{n}(r)\equiv\phi_{n00}(\vec{r})$,
\begin{eqnarray}
\phi_{n}(r)&=&
\pi^{-3/4} b^{-3/2}
\left[L_n^{(1/2)}\left(0\right)\right]^{-1/2}
e^{-r^2/2b^2}
L_n^{(1/2)}\left(r^2/b^2\right),
\label{HOSwf}
\end{eqnarray}
where $L_n^{(\alpha)}$ is the generalized Laguerre polynomial.

The resulting energies (\ref{E}) will depend on
$N_{max}$ as well as $\omega$, $E=E(N_{max},\omega)$.
Since $N_{max}$ is arbitrary, we want the energies not to depend
sensitively on $N_{max}$. This cannot be achieved in general,
but it can for the shallow levels of interest ---that is, those with
$E\simle {\cal O}(1/2\mu R^2)$, which are dominated
by physics at distances $r\simge R$.
As we show in the following,
this is accomplished
by allowing the $C_i^{(\nu)}$ to depend on both 
$N_{max}$ and $\omega$, $C_i^{(\nu)}=C_i^{(\nu)}(N_{max},\omega)$.
Nevertheless, at any order a residual  $N_{max}$ dependence 
introduces an error 
in the calculation of shallow levels,
which should be proportional to powers of $1/\Lambda$.
At the end of the calculation we want to take $N_{max}$
sufficiently large, $\Lambda\simge 1/R$,
so that this error 
is not larger than the error proportional to powers of $R$
stemming from the truncation of Eq. (\ref{Taylorcoord}).

\subsection{LO renormalization}

The physics at LO is obtained by diagonalizing  
the two-body Hamiltonian (\ref{LOpot}). 
The approach is similar to the treatment in a free-particle basis \cite{aleph},
with the difference that
we work here only with bound states, which naturally are closely
related to
HO states because of the presence of the trap. 
Renormalization of the interaction 
at LO has already been discussed in Ref. \cite{trap0}, but for completeness 
we repeat the derivation here. 

We start with the  Schr\"odinger equation (\ref{sch0})
for the wavefunction of the two-fermion system, $\psi^{(0)}(\vec{r})$.
One can show, by inserting Eq. (\ref{expansionS}) into Eq. (\ref{sch0}) 
and projecting  on a HO basis state
$\phi_n(r)$, that the expansion coefficients 
are given by
\begin{equation}
c_n^{(0)}= \kappa^{(0)}
\frac{\phi_n(0)}{E^{(0)}-(2n+3/2) \omega},
\label{cm}
\end{equation}
where  $\kappa^{(0)}=C_0^{(0)}\psi^{(0)}(0)$ is itself a combination 
of the unknown coefficients. 
Direct substitution of the expansion coefficients 
back into Eq. (\ref{expansionS}) 
yields
\begin{eqnarray}
\psi^{(0)}_0(r)&=&
\frac{\kappa^{(0)} \mu}{2 \pi^{3/2}b }
         \exp{(-r^2/2b^2)} 
         \sum_{n=0}^{n_{max}} 
         \frac{L_n^{(1/2)}(r^2/b^2)}
              { \frac{E^{(0)}}{2  \omega} -(n+\frac{3}{4})}.
\label{expansion2}
\end{eqnarray}
{}From the consistency condition at $r=0$
we obtain the relation \cite{trap0} that an energy $E^{(0)}$
has to satisfy:
\begin{eqnarray}
\frac{1}{C_0^{(0)}(n_{max},\omega)} &=& \frac{\mu}{2 \pi^{3/2} b  } 
\sum_{n=0}^{n_{max}} 
\frac{L_n^{(1/2)}(0)}{\frac{E^{(0)}}{2  \omega}- (n+\frac{3}{4})}
\nonumber\\
&=&- \frac{2\mu}{\pi^{2}b}
\left\{
\frac{\Gamma\left(n_{max}+\frac{3}{2}\right)}{\Gamma\left(n_{max}+1\right)}
\left[ 1+ R\left(n_{max}, \frac{E^{(0)}(\omega)}{2\omega}\right) \right]
-\frac{\pi}{2} 
 \frac{\Gamma\left(\frac{3}{4}-\frac{E^{(0)}(\omega)}{2\omega}\right)}
      {\Gamma\left(\frac{1}{4}-\frac{E^{(0)}(\omega)}{2\omega}\right)}
\right\},
\label{C_0_bis}
\end{eqnarray}
where
\begin{eqnarray}
R\left(m, \frac{\varepsilon}{2}\right)
&=&\frac{1-2\varepsilon}
 {8(m+1)\left(m+\frac{7}{4}-\frac{\varepsilon}{2}\right)}
\nonumber\\
&& 
\; _3F_2\left(1, m+\frac{3}{2}, 
              m+\frac{7}{4}-\frac{\varepsilon}{2};
              m+2, m+\frac{11}{4}-\frac{\varepsilon}{2};
              1\right)
\end{eqnarray}
in terms of the generalized hypergeometric function $_3F_2$.
(In Eq. (\ref{C_0_bis}) we used Eqs. (\ref{sumL01}) and (\ref{sumL02}).)
Finally, the constant $\kappa^{(0)}$ is fixed by the chosen normalization of
$\psi^{(0)}_0(r)$. 
(It can be calculated using Eq. (\ref{sumL04}).)
Note that this constant is energy dependent;
when necessary we denote it by $\kappa^{(0)}_{E^{(0)}}$.

The rhs of Eq. (\ref{C_0_bis}) clearly depends on $n_{max}$ and $\omega$,
and so does $C_0^{(0)}$. 
In a given model space, the coupling constant $C_0^{(0)}$ 
can be fixed to 
reproduce one observable, in this case one energy of the two-body system.
If, to be definite, we take that energy as the ground-state energy in the trap,
\begin{equation}
E^{(0)}_0=E_0(\omega),
\end {equation}
then $C_0^{(0)}(n_{max},\omega)$ is determined from
\begin{eqnarray}
\frac{1}{C_0^{(0)}(n_{max},\omega)} 
&=&- \frac{2\mu}{\pi^{2}b}
\left\{
\frac{\Gamma\left(n_{max}+\frac{3}{2}\right)}{\Gamma\left(n_{max}+1\right)}
\left[ 1+ R\left(n_{max}, \frac{E_0(\omega)}{2\omega}\right) \right]
-\frac{\pi}{2} 
 \frac{\Gamma\left(\frac{3}{4}-\frac{E_0(\omega)}{2\omega}\right)}
      {\Gamma\left(\frac{1}{4}-\frac{E_0(\omega)}{2\omega}\right)}
\right\}.
\label{C_0_bisbis}
\end{eqnarray}
While 
one energy in the spectrum 
is fixed in all model spaces, the rest of the energy spectrum 
runs with the model space:
the remaining energies $E^{(0)}_{i\ge 1}=E^{(0)}_{i\ge 1}(n_{max},\omega)$
satisfy Eq. (\ref{C_0_bis}) and in general depend not only
on $\omega$ but also on $n_{max}$.
These energies should converge as $n_{max}\to \infty$ to 
finite values $E^{(0)}_{i\ge 1}(\infty,\omega)$. 
However, once $\Lambda$ in Eq. (\ref{Lambda}) exceeds $1/R$, 
the theoretical errors are dominated by the physics of the effective range 
$r_2$, which was left out of LO.

\subsection{Renormalization beyond LO}

In the two-nucleon system, the effective range is much smaller than the 
scattering length, but is finite. 
In an atomic system near a Feshbach resonance, the range is usually neglected,
although it should become relatively more important 
as one moves away from the resonance.
In either case, the LO, which ignores the range, contains energy-dependent 
errors of ${\cal O}(k^2 a_2 R)$.
In addition, 
in both cases, the truncation to a model space excludes physics of momenta
beyond $\Lambda$, which introduces further energy-dependent errors
of ${\cal O}(k^2 a_2/\Lambda)$.
In other words, the truncation induces contributions to the
effective range, 
the shape parameter, and so on, governed by $\Lambda$ rather than
by $1/R$.
The role of contributions beyond LO is to correct for these 
two types of energy-dependent errors:
NLO for $k^2$ errors
and higher orders for higher powers of $k$.

In the untrapped system, the power counting for a system with a large 
$S$-wave scattering length $a_2$ is such that 
corrections beyond LO have to be treated as perturbations \cite{aleph}. 
It is one of our goals in this paper to show that treating
higher orders in perturbation theory  in the presence
of the trap allows for a systematic improvement of the two-body energies.

At NLO, we include corrections as first-order perturbations
on top of the LO wavefunction $|\psi^{(0)}\rangle$. 
As discussed above, in LO we fix $C_0$ so that one of the states has the 
``observed" energy. The NLO correction in Eq. (\ref{Taylorcoord})
introduces a new parameter $C_2$ that can be chosen
so that a second energy level is fixed. 
However, the NLO term induces, in general, a non-vanishing correction to the 
energy used to fix $C_0$. 
One should, therefore, readjust $C_0$ so that in NLO we reproduce the two 
observables (energy levels) at the same time. In
order to keep track of this change, it is convenient to
split  $C_0$ into an LO piece $C_0^{(0)}$, which remains unchanged,
and an NLO piece $C_0^{(1)}$,
as done in Eq. (\ref{NLOpot}), and treat the latter
in perturbation theory as well. 

Thus, for energies we have
\begin{eqnarray}
E^{(1)}&=&\langle \psi^{(0)}| V^{(1)} |\psi^{(0)}\rangle
=\frac{\kappa^{(0)2}}{C_0^{(0)2}}
\left\{C_0^{(1)}+ 4\mu C_2^{(1)}\left[E^{(0)}
- \frac{C_0^{(0)}}{\pi^{3/2} b^3} \sum_{n=0}^{n_{max}}L_n^{(1/2)}(0)\right]
\right\}
\nonumber\\
&=&
\frac{\kappa^{(0)2}}{C_0^{(0)2}}
\left\{C_0^{(1)}+ 4\mu C_2^{(1)}\left[E^{(0)}
- \frac{4C_0^{(0)}}{3\pi^{2} b^3} 
\frac{\Gamma\left(n_{max}+\frac{5}{2}\right)}
     {\Gamma\left(n_{max}+1\right)}\right]
\right\},
\label{eq:corrNLO}
\end{eqnarray}
where we used Eq. (\ref{sumL03}).

The requirement that two energy levels have the correct positions in the 
spectrum fixes the amount of change from LO, thus providing two equations 
that determine the unknown coupling constants $C_0^{(1)}$ and $C_2^{(1)}$ 
in each model space. 
In the case when the lowest level $E_0$ is already fixed to a given 
(experimental or theoretical) value, $E_0^{(1)}=0$.
However, in the case with finite non-negligible range, 
one can alternatively
choose that in LO $C_0^{(0)}$ be fixed to a level of the two-body 
spectrum without a range, while in NLO that level is shifted to the correct 
position with range, thus requiring that 
$E_0^{(1)}\ne 0$.
These two alternatives are the HO-basis equivalent to
fixing $C_0^{(0)}$ in a free-particle basis to, respectively,
a known binding energy (such as the deuteron
binding energy) or the scattering length.
In either case, if we take, say, the first excited level $E_1(\omega)$ to be
reproduced at NLO in addition to the ground state, 
the two equations for the determination
of $C_0^{(1)}(n_{max}, \omega)$ and $C_2^{(1)}(n_{max}, \omega)$
can be written as
\begin{equation}
E_i^{(1)}(n_{max}, \omega) =E_i(\omega) - E_i^{(0)}(n_{max},\omega),
\qquad i=0,1.
\label{2eqs}
\end{equation}
{}From Eqs. (\ref{eq:corrNLO}) and (\ref{2eqs}), we can easily solve
for $C_0^{(1)}(n_{max}, \omega)$ and $C_2^{(1)}(n_{max}, \omega)$:
\begin{equation}
\frac{4\mu C_2^{(1)}}{C_0^{(0)2}}=
\frac{E_1^{(1)}/\kappa^{(0)2}_{E_1^{(0)}}-E_0^{(1)}/\kappa^{(0)2}_{E_0^{(0)}}}
     {E_1^{(0)}-E_0^{(0)}}
\label{C21}
\end{equation}
and
\begin{equation}
\frac{C_0^{(1)}}{4\mu C_2^{(1)}}=
\frac{4C_0^{(0)}}{3\pi^{2} b^3} 
\frac{\Gamma\left(n_{max}+\frac{5}{2}\right)}
     {\Gamma\left(n_{max}+1\right)}
-\frac{E_0^{(0)} E_1^{(1)}/\kappa^{(0)2}_{E_1^{(0)}}
      -E_1^{(0)} E_0^{(1)}/\kappa^{(0)2}_{E_0^{(0)}}}
      {E_1^{(1)}/\kappa^{(0)2}_{E_1^{(0)}}
      -E_0^{(1)}/\kappa^{(0)2}_{E_0^{(0)}}}.
\label{C01}
\end{equation}
With these coupling constants fixed,
Eq. (\ref{eq:corrNLO}) provides values for the other energy levels.

The form of the NLO wavefunction is a bit more complicated;
in App. \ref{wf_NLO}
we show that, up to higher-order terms,
\begin{eqnarray}
\psi^{(0)}(r)+\psi^{(1)}(r)&=&
\left(1+A^{(1)}\right)  \frac{ \kappa^{(0)} \mu}{ 2 \pi^{3/2}  b} e^{-r^2/2b^2}
\sum_{n=0}^{n_{max}} \frac{L_{n}^{(1/2)}(r^2/b^2)}
     {\frac{1}{2 \omega}(E^{(0)}(n_{max})+E^{(1)}(n_{max}))-(n+\frac{3}{4})}
\nonumber \\
&&+ 2 \mu \kappa^{(0)} \frac{C_2^{(1)}}{C_0^{(0)}} 
\sum_n \phi_n(0)\phi_n(r),
\label{wf2_NLO}
\end{eqnarray} 
where 
\begin{eqnarray}
A^{(1)}=\kappa^{(0)2} \left[ \frac{E^{(1)}}{\pi^{3/2}b^3}    
\sum_m \frac{L_{m}^{(1/2)}(0)}{(E^0-E_m)^3}
+2\mu \frac{C_2^{(1)}}{C_0^{(0)2}} \right].
\label{A1}
\end{eqnarray}

Of course, just as in LO, the levels not used as input at NLO will
have errors $\propto 1/\Lambda$.
As we show explicitly later,
the magnitude of these errors is smaller than at LO,
since more physics
has been accounted for. 
If further precision is desired,
we can continue the procedure to higher orders. 

In this paper we 
consider one order more, N$^2$LO, 
so as to show the systematic trend of improvement
---but we omit details that can be obtained straightforwardly, if painfully.
The correction $E^{(2)}$ to the energy is obtained
using perturbation theory up to the second order.
In addition to the second-order correction 
from $V^{(1)}$ (\ref{NLOpot}), 
one has the first-order correction from $V^{(2)}$ (\ref{NNLOpot}):
\begin{eqnarray}
E^{(2)}= 
\langle \psi^{(0)}| V^{(2)} |\psi^{(0)}\rangle
+\frac{1}{2} \left\{\langle \psi^{(0)}| V^{(1)} |\psi^{(1)}\rangle
+\langle \psi^{(1)}| V^{(1)} |\psi^{(0)}\rangle\right\}.
\label{E2}
\end{eqnarray}
The potential $V^{(2)}$ affects the energy levels that were
fixed already, so again it is convenient to
compensate for this by adding in Eq. (\ref{NNLOpot}) perturbative
shifts $C_0^{(2)}$ and $C_1^{(2)}$ with respect to the lower-order parameters.
These two parameters, together with the four-derivative parameter
$C_4^{(2)}$, are 
determined so that 
three energy levels are fixed to the correct values. 
Taking the lowest three levels $E_i(\omega)$, $i=0,1,2$, to be fixed,
the three equations for the determination
of $C_0^{(2)}(n_{max}, \omega)$, $C_2^{(2)}(n_{max}, \omega)$,
and $C_4^{(2)}(n_{max}, \omega)$ are
\begin{equation}
E_i^{(2)}(n_{max}, \omega) =E_i(\omega) - E_i^{(0)}(n_{max},\omega)
- E_i^{(1)}(n_{max},\omega),
\qquad i=0,1,2.
\label{3eqs}
\end{equation}
Obviously, higher corrections can be added in a similar fashion.

\subsection{Infinite-cutoff limit}
\label{bush_formula_limit}

In the absence of a trap, the LO EFT is formally equivalent \cite{aleph}
 in the infinite-cutoff limit to the pseudopotential \cite{pseudo}.
As we show here, the situation is the same in the presence of
the HO potential, where the pseudopotential was solved in Ref. \cite{HT}.
We consider explicitly
the EFT to NLO, in order to derive also the first corrections
to the pseudopotential in the trap.

The wavefunction to NLO for a finite value of $n_{max}$ 
is given in Eq. (\ref{wf2_NLO}).
By having $n_{max}\to\infty$ and using Eq. (\ref{sumL}) one obtains:
\begin{eqnarray}
\psi(r)&=&-\left(1+A^{(1)}\right) \frac{\kappa^{(0)} \mu}{2\pi^{3/2}  b}
e^{-r^2/2b^2}
\Gamma\left(\frac{3}{4}-\frac{\varepsilon(\infty)}{2}
\right)
U\left(\frac{3}{4}-\frac{ \varepsilon(\infty)}{2},
\frac{3}{2},\frac{r^2}{b^2}\right) \nonumber \\
&&+ 2 \mu \kappa^{(0)} \frac{C_2^{(1)}}{C_0^{(0)}}  \delta (\vec{r}).
\label{wf_NLO_inf}
\end{eqnarray}
where $U$ is the confluent hypergeometric function, 
and we have introduced the energy
\begin{equation}
\varepsilon(\infty)=\frac{E^{(0)}(\infty, \omega)+E^{(1)}(\infty, \omega)}
                         {\omega},
\end{equation}
which is the limit of the energy (in units of $\omega$) of the 
two-body system in the trap.
The second term in Eq. (\ref{wf_NLO_inf}) was obtained by using the 
completeness 
of the HO basis,
\begin{eqnarray}
\lim_{n_{max}\to\infty}   \sum_{n=0}^{n_{max}} \phi_n(0)\phi_n(r)
=\lim_{N_{max}\to\infty} \sum_{nl=0}^{N_{max}}\sum_{m=-l}^{l}
\langle 0 |nlm\rangle \langle nlm|\vec{r}\rangle
=\delta (\vec{r}).
\end{eqnarray}
The singularity of this term is mitigated by the pre-factor
$C_2^{(1)}/C_0^{(0)}$, which vanishes as $1/\Lambda$ for large cutoff.
It stems from the enhancement of high virtual momenta due to
the second derivatives in $V^{(1)}$.
As shown in App. \ref{AppD}, for an equivalent 
energy-dependent potential this term is absent.

For small, non-vanishing values of $r$, {\it i.e.} $0<r\ll b$, 
use of Eq. (\ref{smallx}) gives
\begin{equation}
\psi(0<r\ll b)
\propto\frac{1}{r} \left\{1
-2\frac{\Gamma\left(\frac{3}{4}-\frac{\varepsilon(\infty)}{2}\right)}
       {\Gamma\left(\frac{1}{4}-\frac{\varepsilon(\infty)}{2}\right)} 
  \frac{r}{b} 
 + {\cal O}\left(\frac{r^2}{b^2}\right)\right\}.
\label{expansion_NLO}
\end{equation}
By identification of this wavefunction
with the wavefunction in the untrapped case, Eq. (\ref{aswf}),
we obtain the relation between the energy of the trapped two-body system 
and the ERE parameters:
\begin{eqnarray}
\frac{\Gamma\left(\frac{3}{4}-\frac{\varepsilon(\infty)}{2}\right)}
      {\Gamma\left(\frac{1}{4}-\frac{\varepsilon(\infty)}{2}\right)} 
=\frac{b}{2 a_2} \left\{1-\frac{a_2 r_2}{b^2} \varepsilon(\infty)+\ldots 
               \right\}.
\label{transcend}
\end{eqnarray}

In our framework, Eq. (\ref{transcend}) is to be interpreted in perturbation
theory.
At LO only the scattering length is taken into account, 
and the energies $\varepsilon^{(0)}(\infty)$ are given by 
\begin{eqnarray}
\frac{\Gamma\left(\frac{3}{4}-\frac{\varepsilon^{(0)}(\infty)}{2}\right)}
      {\Gamma\left(\frac{1}{4}-\frac{\varepsilon^{(0)}(\infty)}{2}\right)} 
=\frac{b}{2a_2},
\label{transcendLO}
\end{eqnarray}
a relation first found in Ref. \cite{HT} using the 
pseudopotential \cite{pseudo}. The latter can be viewed as a renormalization
of the delta-function interaction \cite{aleph}.
Indeed, the strength of the LO delta-function interaction has to decrease 
with the increase
of the model-space size in order for it to give sensible results.
One can see
from Eq. (\ref{asyGr})
that for large $n_{max}$ the 
first term  
in Eq. (\ref{C_0_bisbis}) grows as $\sqrt{n_{max}}$.
Taking the limit of Eq. (\ref{C_0_bisbis}) and using Eq. (\ref{transcendLO}),
one finds
\begin{eqnarray}
C_0^{(0)}&=&
- \frac{\pi^{2}}{\mu  \Lambda}  \left[ 1
+ \frac{\pi}{2\Lambda a_2} 
+{\cal O}\left(\frac{1}{\Lambda^2b^2}
\right)\right].
\label{C_0run}
\end{eqnarray}
In the $\Lambda b \to \infty$ limit this running is exactly the
one found in free space \cite{aleph}:
the non-trivial rate of change with the cutoff $\Lambda$ is controlled by 
$1/a_2$.
This is not surprising since the large-$\Lambda$ behavior 
should be independent of the long-range physics of the trap.

At NLO, the effective range appears: Eq. (\ref{transcend}) becomes
\begin{eqnarray}
\frac{\Gamma\left(\frac{3}{4}-\frac{\varepsilon^{(0)}(\infty)
+\varepsilon^{(1)}(\infty)}{2}\right)}
      {\Gamma\left(\frac{1}{4}-\frac{\varepsilon^{(0)}(\infty)
+\varepsilon^{(1)}(\infty)}{2}\right)} 
=\frac{b}{2 a_2} \left\{1-\frac{a_2 r_2}{b^2} \varepsilon^{(0)}(\infty)
                 \right\},
\label{transcendNLO}
\end{eqnarray}
which can be solved to this order as
\begin{equation}
\varepsilon^{(1)}(\infty)= \frac{2 a_2 r_2}{b^2} 
\frac{\varepsilon^{(0)}(\infty)}
     {\psi^{(0)}\left(\frac{3}{4}-\frac{\varepsilon^{(0)}(\infty)}{2}\right)
     -\psi^{(0)}\left(\frac{1}{4}-\frac{\varepsilon^{(0)}(\infty)}{2}\right)},
\label{energyexpansion_NLO}
\end{equation}
in terms of the digamma function $\psi^{(0)}$. 
In the presence of range, all levels change from LO to NLO.
The range in fact controls the asymptotic behavior
of the NLO coupling constants.
In the large-$n_{max}$ limit we find from Eqs. (\ref{C01}) and (\ref{C21})
that
\begin{equation}
C_0^{(1)}=-\frac{\pi^3 r_2}{12\mu} 
\left[1+{\cal O}\left(\frac{1}{r_2 \Lambda}, \frac{1}{a_2 \Lambda}\right)
\right]
\label{C01asym}
\end{equation}
and
\begin{equation}
\frac{C_2^{(1)}}{C_0^{(0)2}}= \frac{\mu r_2}{8\pi} 
\left[1+{\cal O}\left(\frac{1}{r_2 \Lambda}\right)\right].
\label{C21asym}
\end{equation}
Again, this is the same running as in free space \cite{aleph},
as it should be.
Note that it qualitatively changes for $r_2=0$.
In this case, $\varepsilon^{(1)}(\infty)=0$,
which means, given our choice of levels to fix the coupling constants
at NLO, that
$E_0^{(1)}=0$ and $E_1^{(1)}={\cal O}(E_1^2/\Lambda)$.
As a consequence the ratio of energies in Eq. (\ref{C21})
goes to $0$ as $\Lambda\to \infty$,
and $\mu \Lambda C_0^{(1)}$ and
$\mu \Lambda^3 C_2^{(1)}$ approach constants. 

At N$^2$LO the power counting of Ref. \cite{aleph} indicates
that no new ERE term should be included. This is a reflection of
the fact that a fine-tuning in $a_2$ does not in general lead
to an enhancement in the shape parameter $P_2$:
in order to go from $r_2 k^2/2$ to $P_2k^4/4$ two orders are needed.
Therefore at N$^2$LO Eq. (\ref{transcend}) is unmodified.
It is only at N$^3$LO that we need to account for a non-vanishing 
$S$-wave shape parameter, also with an $S$-wave potential
of the form of Eq. (\ref{NNLOpot}) (but with different parameters).
For fermions, N$^3$LO also contains a $P$-wave interaction 
to account for the $P$-wave scattering volume. 

It is clear that the procedure can be continued to higher orders,
and we expect for $l=0$ levels
\begin{eqnarray}
\frac{\Gamma\left(\frac{3}{4}-\frac{\varepsilon(\infty)}{2}\right)}
      {\Gamma\left(\frac{1}{4}-\frac{\varepsilon(\infty)}{2}\right)} 
=-\frac{b k}{2} \cot\delta_0(k),
\label{transcendallO}
\end{eqnarray}
where 
\begin{equation}
kb=\sqrt{2\varepsilon(\infty)}
\end{equation} 
and 
$\delta_0(k)$ is given by the ERE (\ref{ERE}).
This extension to subleading orders 
agrees with Refs. \cite{models,DFT_short,mehen}. 
Waves with $l\ge 1$ can also be examined with the same method we 
developed here, but we leave details for future work.

The importance of Eq.~(\ref{transcend}) lies in the link between
the energies inside the HO well, $\varepsilon(\infty)$,
and the scattering parameters, $a_2$, $r_2$, {\it etc.}
It is the analog of L\"uscher's formula \cite{luescher}, which links the
levels inside a cubic box and the same scattering parameters.
As such,
Eq. (\ref{transcend}) provides the energy levels necessary to fix the 
coupling constants in finite model spaces when the ERE parameters
are known.
In the nuclear case, for example, the scattering parameters have been 
determined from data, so Eq. (\ref{transcend}) can be used to fix
the parameters of the pionless EFT without relying on a fit
to nuclear levels,
as done in Ref. \cite{NCSM}.
An extension of Ref. \cite{NCSM} to this case is in progress.

In the atomic case, the lowest energy $E_0 (\omega)$ of two trapped particles 
has been measured \cite{kohl}. One can use $E_0 (\omega)$ directly as input
in Eq. (\ref{C_0_bisbis}).
Alternatively, 
this energy was found \cite{kohl}  to be in good agreement with the 
lowest state of the theoretical spectrum obtained under a pseudopotential 
assumption \cite{HT}, in which the eigenvalues are determined by the scattering
length $a_2$. 
In Ref. \cite{trap0} we used this theoretical LO energy as input;
we confirmed that
the two-fermion spectrum of the underlying pseudopotential is reached 
asymptotically as $n_{max}\to\infty$, 
and we 
calculated the properties
of three- and four-fermion systems.
In subleading orders, we can now include effective range and higher 
effects,
which might account for the
small discrepancies \cite{kohl} between the theory and experiment.

We should stress that, because the bare coupling constants are not observables,
their sizes have no direct physical meaning. 
What matters is the total contribution of a given order to an observable.
For example,
for $\Lambda\simge 1/r_2$, $C_0^{(0)}\simle C_0^{(1)}$,
and yet, the NLO energy shift (\ref{energyexpansion_NLO})
is small as long as $r_2$ is sufficiently small.
We show explicit results for energies in the next section.

\section {Results}
\label{sect3}

In this section, we illustrate the approach described above in a few cases of 
interest for atomic and nuclear systems. We initially consider situations 
where the range of the interaction can be neglected;
we first look at the unitarity limit, $b/a_2=0$, and then finite $b/a_2$.
Finally, we consider the case  
where the range of the interaction is non-negligible {\it albeit} still 
small with respect to the scattering length. 
In all cases, we use Eq. (\ref{transcend}) to provide
the asymptotic levels in the trap.

\subsection{Unitarity limit}

We start by considering a system of two trapped particles at unitarity, 
characterized by $b/a_2=0$ and $r_2/b=0$. 
In the untrapped case,
this situation can be realized by considering an attractive potential 
given by, for instance, a  square well with a fine-tuned depth. 
Asymptotically, the wavefunction at zero energy behaves as $1/r$  
and the scattering length $a_2$ is then infinite.

In the presence of the harmonic trap, the only scale is set by $b$, or, 
equivalently, $\omega$,
so the solutions of Eq. (\ref{transcend}) have to be constants;
indeed, they 
are given by the poles of the Gamma functions in the denominators, 
\textit{i.e.,}
\begin{eqnarray}
\varepsilon_n(\infty) =\frac{1}{2}+ 2n, 
\label{trans_unit}
\end{eqnarray}
where 
$n\geq 0$ is an integer.
At each order we use a finite number of these energies
to determine the interaction parameters in each model space:
$\varepsilon_0$ at LO, $\varepsilon_0$ and $\varepsilon_1$ at NLO,
and $\varepsilon_0$, $\varepsilon_1$, and $\varepsilon_2$ at N$^2$LO.

\begin{figure}[tb]
\begin{center}
\includegraphics*[scale=0.80,angle=-90]{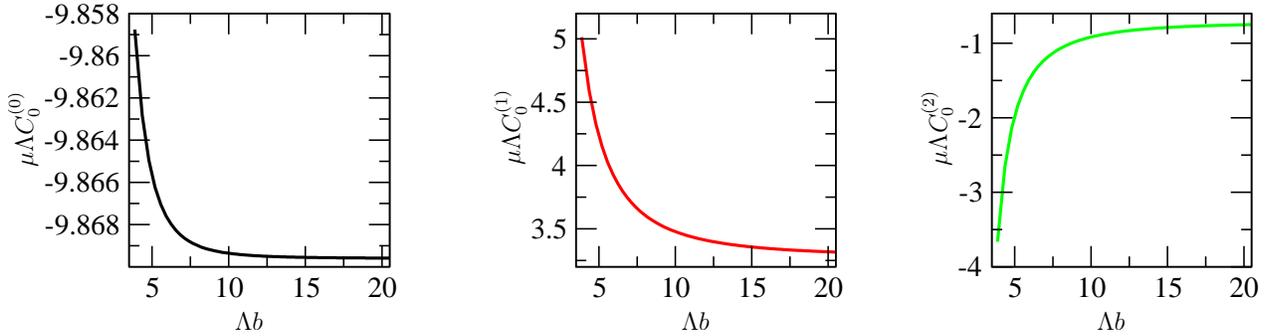}
\end{center}
\caption{The coupling constants $\mu \Lambda C_0^{(0)}$, 
$\mu \Lambda C_0^{(1)}$, and $\mu \Lambda C_0^{(2)}$ 
at unitarity as a function of the cutoff in the dimensionless 
combination $\Lambda b$.}
\label{C0_lambda_unitarity}
\end{figure}

\begin{figure}[tb]
\vspace*{2.1 cm}
\begin{center}
\includegraphics*[scale=0.60,angle=-90]{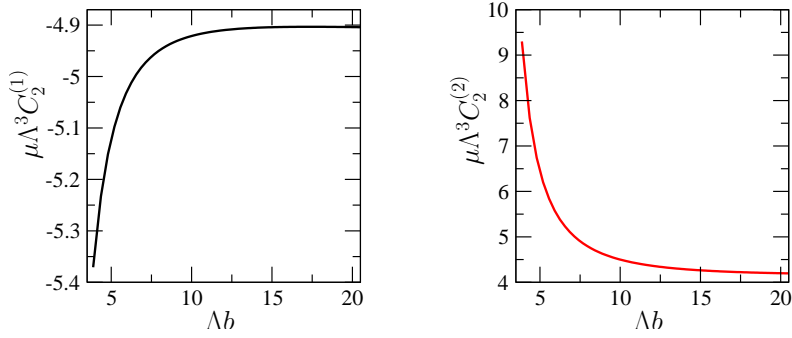}
\end{center}
\caption{The coupling constants $\mu \Lambda^3 C_{2}^{(1)}$
and $\mu \Lambda^3 C_{2}^{(2)}$ at unitarity 
as a function of the cutoff in the dimensionless combination $\Lambda b$.}
\label{C2_lambda3_unitarity}
\end{figure}

\begin{figure}[tb]
\begin{center}
\includegraphics*[scale=0.4,angle=-90]{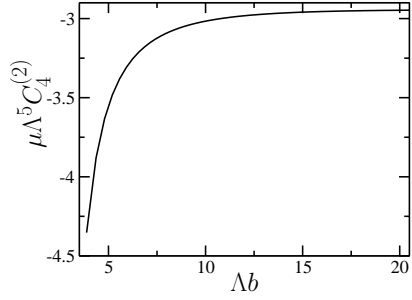}
\end{center}
\caption{The coupling constant $\mu \Lambda^5 C_{4}^{(2)}$ at unitarity 
as function of the 
cutoff in the dimensionless combination $\Lambda b$.}
\label{C4_lambda5_unitarity}
\end{figure}

Using the ground-state energy to determine $C_0^{(0)}$, 
Eq. (\ref{C_0_bisbis}) simplifies to
\begin{equation}
C_0^{(0)}(n_{max},\omega)=
- \frac{\pi^{2}b}{2\mu}
\frac{\Gamma\left(n_{max}+1\right)}{\Gamma\left(n_{max}+\frac{3}{2}\right)},
\label{C_0unit}
\end{equation}
which at large cutoff becomes
\begin{equation}
\mu  \Lambda C_0^{(0)}=
- \pi^{2}\left[ 1+ {\cal O}\left(\frac{1}{\Lambda^4 b^4}\right)\right].
\label{C_0rununit}
\end{equation}
Analogous expressions can be derived for the other coupling constants.
The resulting running of the coupling constants is shown 
in Figs. \ref{C0_lambda_unitarity}, \ref{C2_lambda3_unitarity},
and \ref{C4_lambda5_unitarity}. 
One can see that, in agreement with Sec. \ref{bush_formula_limit},
a coupling constant $C_i$ behaves asymptotically as 
$1/\Lambda^{2i+1}$.

\begin{figure}[tb]
\begin{center}
\includegraphics*[scale=0.82]{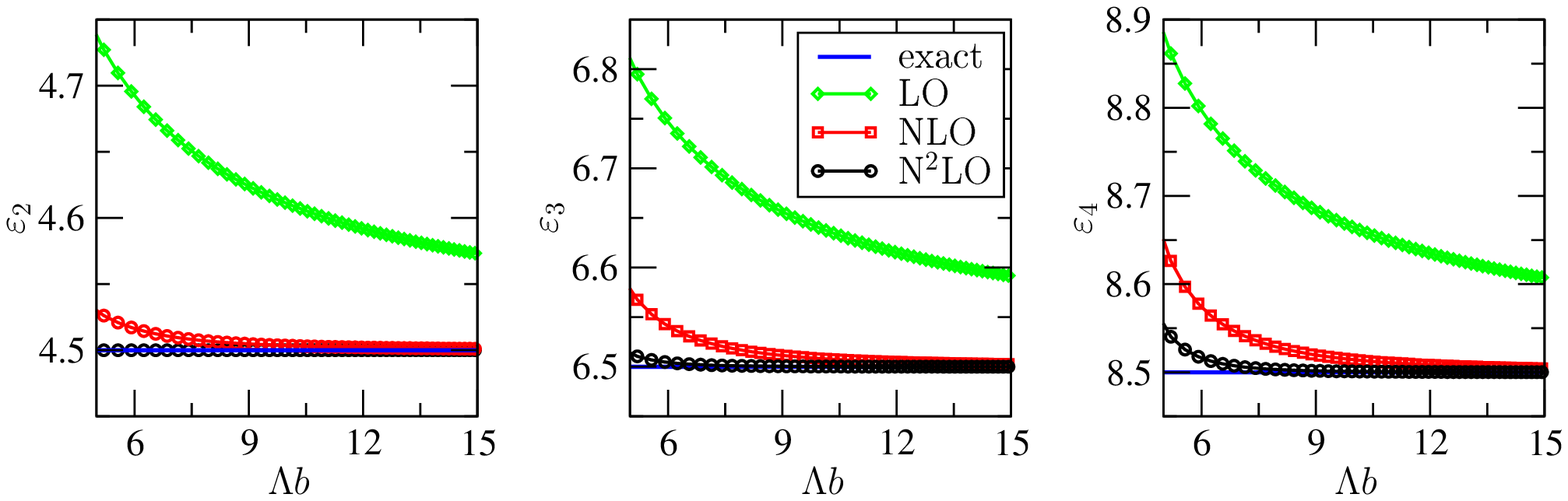}
\end{center}
\caption{Running with the dimensionless cutoff $\Lambda b$
of the second ($\varepsilon_2$), third ($\varepsilon_3$), and fourth 
($\varepsilon_4$) excited-state energies, 
in units of $\omega$, 
for two particles in a trap with the interaction characterized by $b/a_2=0$ 
and $r_2/b=0$. 
Results at LO (green diamonds), NLO (red squares), and  N$^2$LO (black circles)
are compared with the exact values (blue solid lines) 
given by Eq. (\ref{transcend}).
The coupling constants are fixed so that the ground-state, 
additionally the first excited-state,
and additionally the second excited-state energies 
are reproduced in LO, NLO, and N$^2$LO, respectively. 
(Note that, indeed, at N$^2$LO the second excited state in the leftmost panel
is constant on top of 
the exact value for all values of $\Lambda b$.)}
\label{states_unitarity}
\end{figure}

With the interaction parameters so determined,
we calculate
the remaining energies, which depend on $n_{max}$. 
Results for some excited states, from the second ($n=2$)
to the fourth ($n=4$), are shown in 
Fig. \ref{states_unitarity} for different values of the cutoff
$n_{max}$ in the dimensionless combination $\Lambda b$. 
At LO and NLO, all states change with $\Lambda b$; 
at N$^2$LO, the second excited-state energy is used as input.
As we showed in Sec. \ref{bush_formula_limit}, these calculated energies
converge as $n_{max}$ increases to the values given in Eq. (\ref{trans_unit}).
The plot explicitly shows, in addition, that
the convergence with respect to $n_{max}$ to the exact value 
is increased as higher-order corrections are added in perturbation theory.

\begin{figure}[tb]
\begin{center}
\includegraphics*{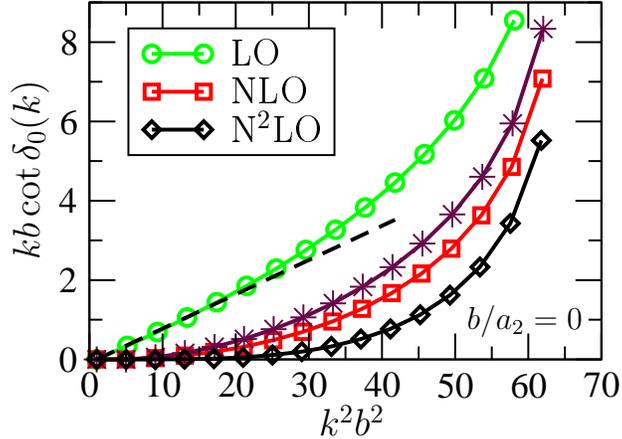}
\end{center}
\caption{$S$-wave scattering phase shifts $k b\cot\delta_0(k)$
as function of the dimensionless squared momentum $k^2b^2$
in the finite model space characterized by $n_{max}=15$,
at unitarity.
The points at LO (green circles), NLO (red squares),
and N$^2$LO (black diamonds) 
are obtained from calculated energies via Eq. (\ref{transcendallOinv}).
The dashed line 
corresponds to a linear fit of the LO curve at small $kb$ values.
For comparison, we also show results with the NLO potential  
fully diagonalized (black stars).}
\label{phase_shift_unitarity_all_diag}
\end{figure}

In order to further study the effects of a finite $n_{max}$,
we consider Eq. (\ref{transcend}) from a different angle. 
Hitherto, we have used it simply to fix observables (energy levels). 
Now we employ it also to extract phase shifts.
In a given model space, characterized by a given $n_{max}$,
we calculate the energy levels $\varepsilon_n(n_{max})$.
The lowest levels used in the fitting of coupling constants
are, of course, exact, while all others deviate from the exact values.
The momentum $k_n$ 
associated with an energy level $\varepsilon_n(n_{max})$ is
\begin{equation}
k_n b=\sqrt{2\varepsilon_n(n_{max})},
\end{equation}
so that the phase shifts for discrete values of $k$,
$\delta_0(k_n)$, can be determined by 
simply inverting Eq. (\ref{transcendallO}):
\begin{equation}
k_nb\cot\delta_0(k_n)=-2\frac{\Gamma(3/4-\varepsilon_n(n_{max})/2)}
                             {\Gamma(1/4-\varepsilon_n(n_{max})/2)}.
\label{transcendallOinv}
\end{equation}
This allows us to study the effect of the truncation in a HO basis in terms 
of the more familiar ERE parameters: 
any deviation from the value $kb \cot\delta_0(k)=0$ is due to truncation errors
and we can calculate the induced range, shape parameter, and so on. 
In Fig. \ref{phase_shift_unitarity_all_diag}, we plot $k b\cot\delta_0(k)$ 
in a model space with $n_{max}=15$ as a function of $k^2b^2$. 
(It is more natural to express $k$ in units of 
$1/a_2$, but at unitarity $b$ provides the only length unit.) 
At LO, $kb \cot \delta_0(k)$ starts off linear in $k^2b^2$:
a linear fit (indicated by the dashed line) shows
an induced effective range of about $0.17b$ in this 
particular model space.
At larger values of $k^2b^2$ deviation from the linear
behavior is seen, indicating the presence of further induced ERE
parameters.
NLO and N$^2$LO corrections reduce the size of the ERE parameters, so that the 
results for the phase shifts improve order by order, getting closer and closer 
to the horizontal axis. Since here $\Lambda^2b^2= 63$, at $k^2b^2\sim 60$, 
the errors are dominated by the higher orders, and therefore the lower orders 
do little to improve on the previous orders. However, at low momentum, the 
results systematically improve, as expected. 

Our perturbative treatment thus provides a way to systematically
reduce the effect of truncation to a reduced model space,
demanded in the calculation of larger systems.
Emboldened by this success, 
one might be tempted to treat the 
subleading potentials exactly in the 
Schr\"odinger equation. 
Apart from the renormalization
problems pointed out in Ref. \cite{aleph},
we find no obvious numerical improvement when this is done here. 
In Fig. \ref{phase_shift_unitarity_all_diag} we also show for comparison 
results obtained when the NLO correction to the potential
is not considered as a perturbation but fully diagonalized together
with the LO potential. 
(The same two lowest levels are used to fix the two
coupling constants of the LO+NLO potential.)
One can see that by considering the LO+NLO potential 
this way, results
are further away from the exact curve than by treating the
NLO potential as a perturbation. 
This result is not particular to this example; 
more generally, truncation errors get worse when
subleading corrections are treated improperly. 
The reason is that doing so includes only part of the higher-order
corrections; it neglects the rest, needed to ensure systematic improvement.

\subsection{Finite scattering length and zero range}

In the previous subsection we saw that the rate of approach to the 
asymptotic
values of two-body energies improves systematically as the order increases
at unitarity.
We now show that there are no qualitative changes when we
consider the case of a pseudopotential
away from unitarity \cite{HT},
where the scattering length $a_2$ is finite,
with $r_2$ still vanishing. 

The corrections to the potential are taken into account
as in the unitarity case, {\it i.e.}, the LO correction is iterated 
to all orders whereas higher corrections are treated as perturbations.
The parameters at each order are adjusted so that
the lowest levels satisfy Eq. (\ref{transcend}) exactly. 
The $\Lambda$ dependence of coupling constants is similar
to unitarity, except for a markedly slower convergence for $C_0$.
This is particularly obvious for $C_0^{(0)}$, when we
compare the more general Eq. (\ref{C_0run}),
which applies here, with its unitarity version (\ref{C_0rununit}).

\begin{figure}[tb]
\begin{center}
\includegraphics*[scale=0.82]{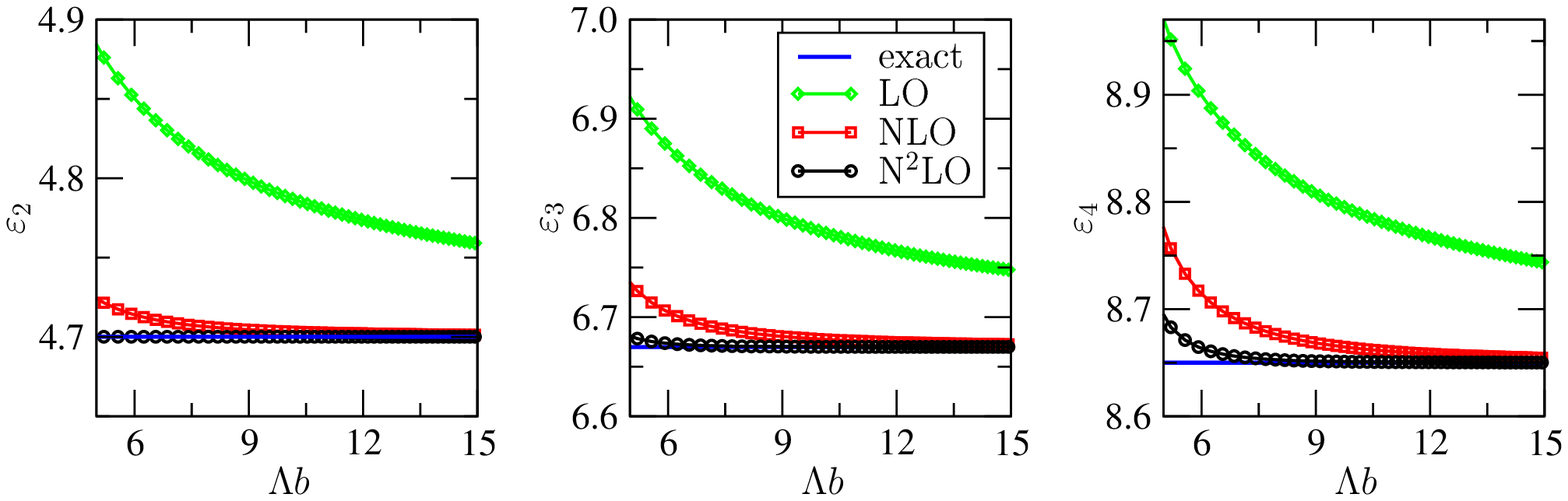}
\end{center}
\caption{Same as in Fig. \ref{states_unitarity}, but for $b/a_2=-1$.
}
\label{states_b_over_a_min1}
\end{figure}

As for energies, let us first consider $a_2<0$.
This situation corresponds
to the case when the depth of the potential between the two particles 
is decreased starting from the fine-tuned value at unitarity. 
As an example,
we show in  Fig. \ref{states_b_over_a_min1}
results for $b/a_2= -1$
for the same energy levels previously displayed at unitarity
in Fig. \ref{states_unitarity}.
The exact values are slightly higher than at unitarity.
As before, convergence of the energies to the exact values is improved as 
more corrections to the potential 
are added, and the difference between the truncated-space energy and 
the exact result is mitigated as more corrections are included.

For a weak enough potential, $|a_2|$ becomes small. 
The energies are then close to the HO energies,
\begin{equation}
\varepsilon_n= \frac{3}{2} +2n + \delta\varepsilon_n ,
\label{eigv_loI}
\end{equation} 
where $\delta\varepsilon_n$ is a small correction and $n=0,1,\ldots$. 
By using Eq. (\ref{transcend}) and expanding the Gamma function around 
its poles one finds
\begin{equation}
\delta\varepsilon_n(\omega)= -\frac{4}{\Gamma\left(-\frac{1}{2}-n\right) P_n} 
\frac{a_2}{b} \left[1+{\cal O}\left(\frac{a_2}{b}\right)\right],
\label{deltaeigv_loI}
\end{equation}
where
\begin{equation}
 P_n = \lim_{z\to -n} \left[(z+n) \Gamma(z) \right]^{-1} 
= \left ( 1+\frac{1}{n}\right )^n\prod_{m=1, m\ne n}^{\infty}
            \left(1+\frac{1}{m}\right)^n \left(1-\frac{n}{m}\right).
\label{Pn}
\end{equation}
We have verified numerically that 
in this case convergence to the exact value can even be sped up by 
considering {\it all} interactions as perturbations, 
as in the ``natural'' continuum case discussed in Ref. \cite{aleph}.

\begin{figure}[tb]
\begin{center}
\includegraphics*[scale=0.82]{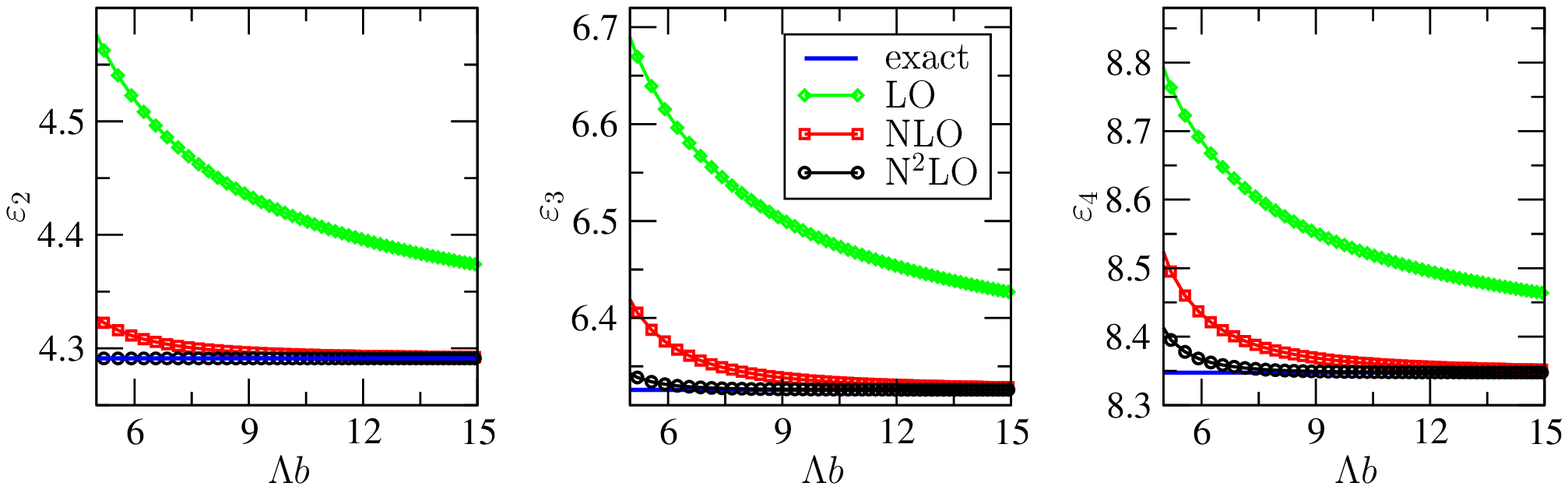}
\end{center}
\caption{Same as in Fig. \ref{states_unitarity}, but for $b/a_2=1$.}
\label{states_b_over_a_1}
\end{figure}

We now turn to the case where $a_2>0$. 
As the interaction between particles becomes stronger 
(starting from the case at unitarity)
the scattering length $a_2$ decreases. 
The ground state can have negative energy.
The plot of the energy for excited states (starting from the second) 
for $b/a_2=1$ is shown in Fig. \ref{states_b_over_a_1}. 
The exact values are now slightly lower than at unitarity.
Again, there is no qualitative change in the pattern of convergence
with respect to unitarity or negative $a_2$.

For a strong enough interaction, the absolute value of the ground-state energy,
$|\varepsilon_0(\infty)|$,  
is large and in the trapped system we have
\begin{equation}
\frac{\Gamma(3/4-\varepsilon_0(\infty)/2)}
     {\Gamma(1/4-\varepsilon_0(\infty)/2)}
= \sqrt{-\frac{\varepsilon_0(\infty)}{2}} 
  \left[1+{\cal O}\left(\varepsilon_0^{-2}(\infty)\right)\right].
\end{equation}
Using Eq. (\ref{transcend}), we find
\begin{equation}
\varepsilon_0(\infty)= -\frac{1}{2} \frac{b^2}{a_2^2}
                \left[1+{\cal O}\left(\frac{a_2^4}{b^4}\right)\right].
\label{deutBE}
\end{equation}
Thus, when $b/a_2\to \infty$, the ground-state energy 
$E_0(n_{max},\omega)$ converges to
$E_0(\infty)=-1/2\mu a_2^2$, the untrapped bound-state energy 
for the case $a_2\gg R$.
The excited levels are again close to the HO values
\begin{equation}
\varepsilon_n= \frac{3}{2} +2(n-1) + \delta\varepsilon_n
=-\frac{1}{2} +2n + \delta\varepsilon_n,
\label{eigv_loIII}
\end{equation}
where $n=1,2,\ldots$.  
Following the same procedure as for large, negative $b/a_2$, we obtain
the corrections
\begin{equation}
\delta\varepsilon_{n\ge 1}= -\frac{4}{\Gamma\left(\frac{1}{2}-n\right) P_{n-1}}
 \frac{a_2}{b} \left[1+{\cal O}\left(\frac{a_2}{b}\right)\right].
\label{deltaeigv_loIII}
\end{equation}

\subsection{Interaction with finite range}

A finite interaction range $R$ usually generates higher ERE parameters
of the same magnitude, $|r_2|\sim R$, $|P_2| \sim R^3$, {\it etc.},
even when there is fine-tuning that leads to $|a_2|\gg R$.
As a last example, we account for a finite effective range $r_2$.

Regardless of the quantity chosen as input in LO, 
the existence of range introduces errors that are energy dependent and can
only be accounted for in subleading orders.
As before, we use as LO input the ground-state energy
$\varepsilon_0$ given by Eq. (\ref{transcendLO}). 
This fixes the 
running of $C_0^{(0)}$ to the same values as in the previous subsection.
However, at NLO, we obtain $C_0^{(1)}$ and $C_2^{(1)}$
from the first two states 
of Eq. (\ref{transcend}) with $a_2$ {\it and} $r_2$ non-vanishing.
NLO is, thus, different from the previous subsection: 
it accounts
not only for errors ${\cal O}(a_2 k^2/\Lambda)$ due to the explicit 
truncation to the model space but
also for implicit ones, ${\cal O}(a_2 R k^2)$, in the potential.

As discussed in Sec. \ref{bush_formula_limit}, the N$^2$LO corrections are 
slightly more subtle \cite{aleph}. 
The introduction of range leaves an error that can be as big as
${\cal O}(a_2^2 r_2^2 k^4)={\cal O}(a_2^2 R^2 k^4)$.
Errors from the explicit truncation of the model space are now
${\cal O}(a_2^2 k^4/\Lambda^2)$ or ${\cal O}(a_2^2 r_2 k^4/\Lambda)$,
and, as in general, are smaller than errors from the truncation of
the expansion once $\Lambda \simge 1/R$.
These types of errors are one order
in $kR$ or $k/\Lambda$ from NLO, which requires $V^{(2)}$ for control.
In contrast, errors from the shape parameter are only 
${\cal O}(a_2 P^3 k^4)={\cal O}(a_2 R^3 k^4)$, 
two orders in $kR$ down from NLO.
Thus, at N$^2$LO we determine $C_0^{(2)}$, $C_2^{(2)}$, and $C_4^{(2)}$
from the lowest three levels of Eq. (\ref{transcend}),
still with non-vanishing $a_2$ and $r_2$ and neglecting
all higher-order ERE parameters.

\begin{figure}[tb]
\vspace*{2.4 cm}
\begin{center}
\includegraphics*[scale=0.80,angle=-90]{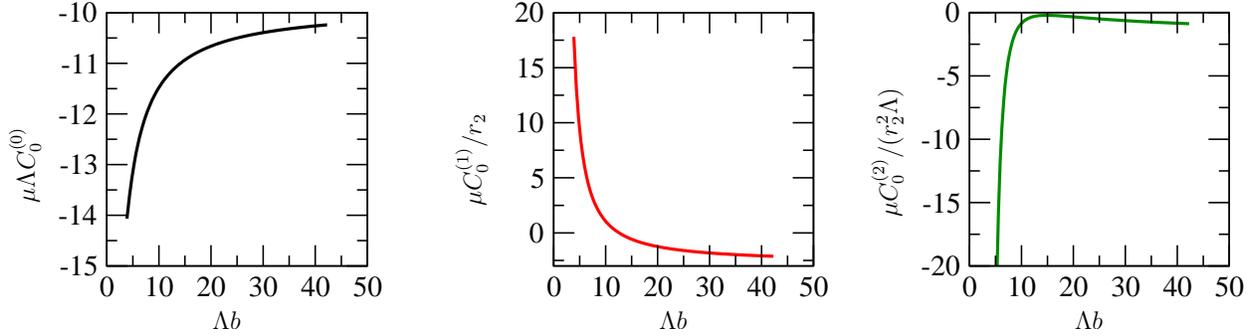}
\end{center}
\caption{The coupling constants 
$\mu \Lambda C_0^{(0)}$, $\mu C_0^{(1)}/r_2$, and $\mu C_0^{(2)}/r_2^2\Lambda$ 
for $b/a_2=1$ and $r_2/b= 0.1$ as function of
the cutoff in the dimensionless combination $\Lambda b$.}
\label{C0_lambda_b_a_1}
\end{figure}

\begin{figure}[tb]
\vspace*{2.1 cm}
\begin{center}
\includegraphics*[scale=0.60,angle=-90]{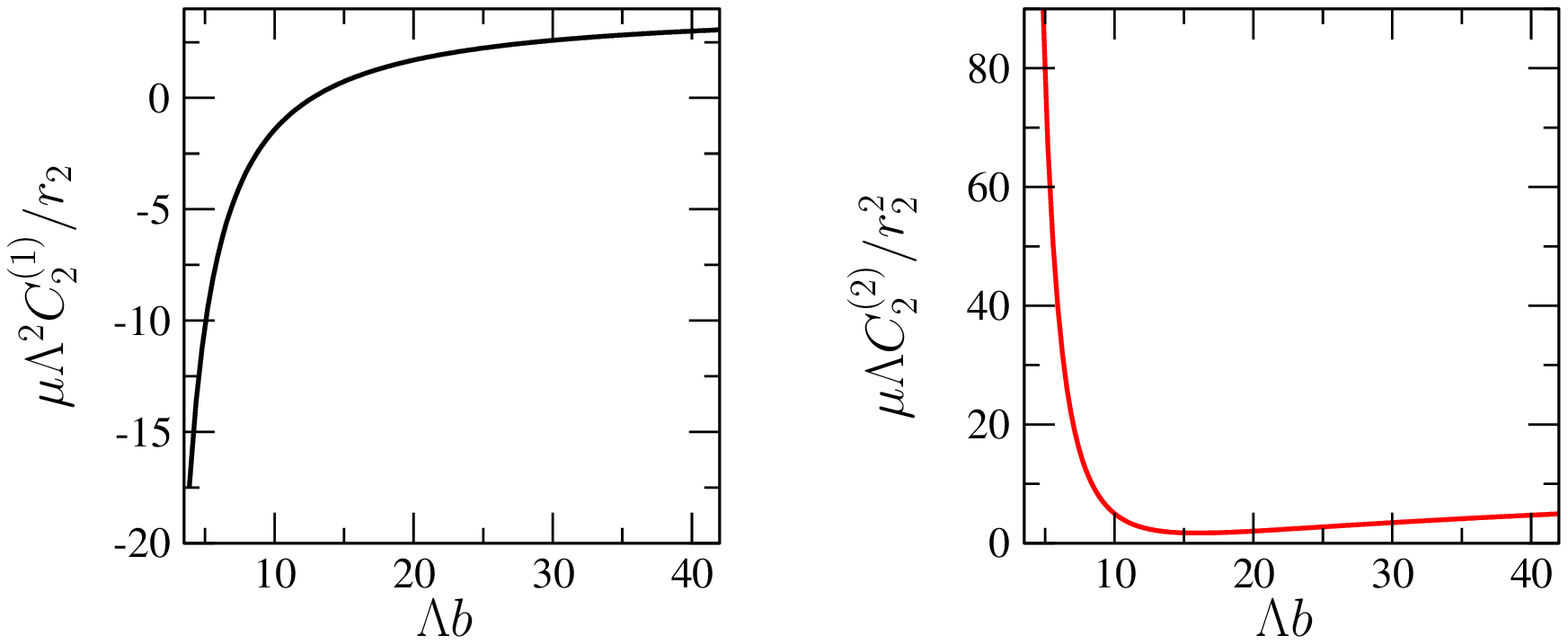}
\end{center}
\caption{The coupling constant 
$\mu \Lambda^2 C_{2}^{(1)}/r_2$ and $\mu \Lambda C_{2}^{(2)}/r_2^2$ 
for $b/a_2=1$ and $r_2/b= 0.1$ 
as a function of the cutoff in the dimensionless combination $\Lambda b$.}
\label{C2_lambda3_b_a_1}
\end{figure}

\begin{figure}[tb]
\vspace*{2.1 cm}
\begin{center}
\includegraphics*[scale=0.4,angle=-90]{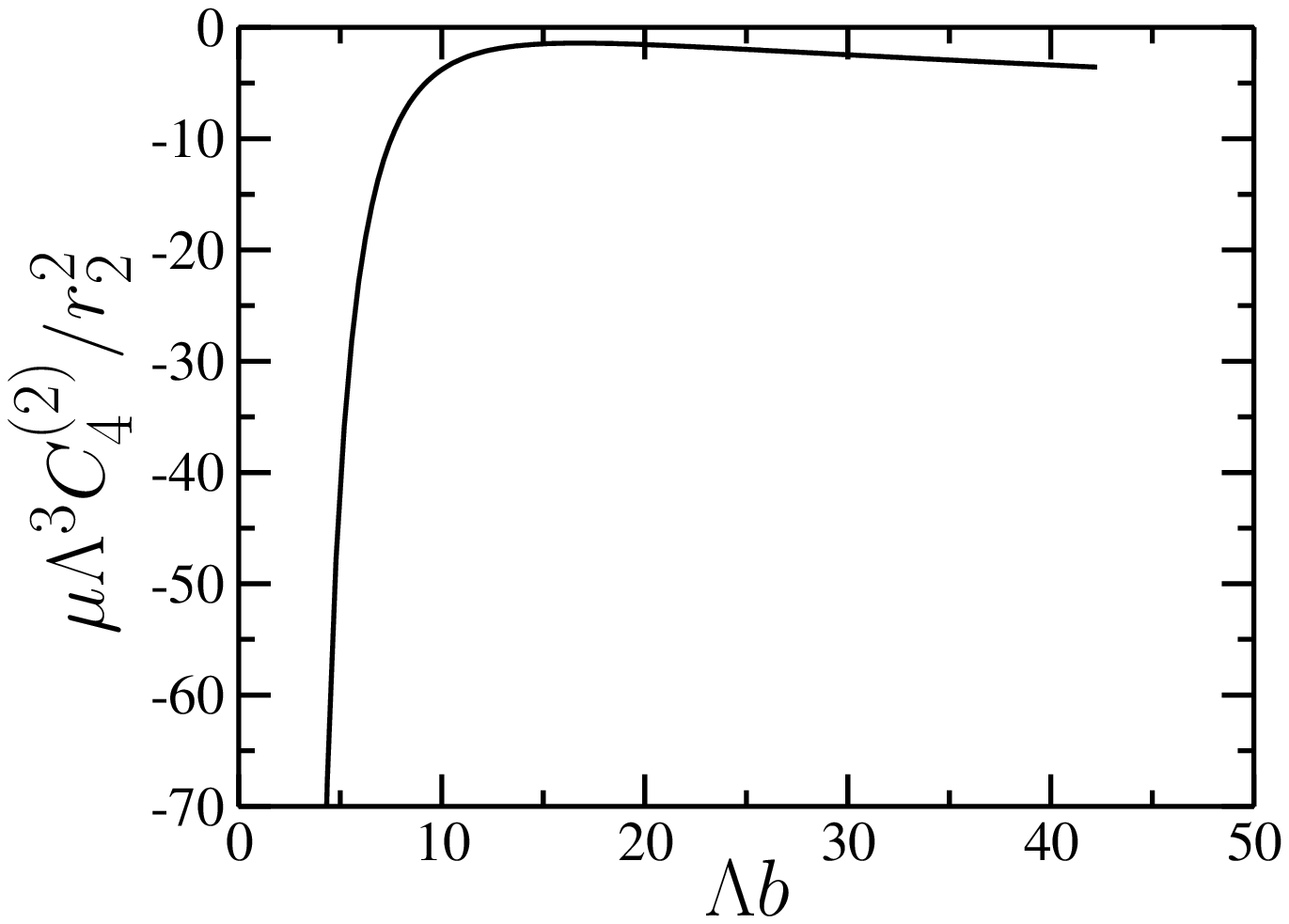}
\end{center}
\caption{The coupling constant 
$\mu \Lambda^3 C_{4}^{(2)}/r_2^2$
for $b/a_2=1$ and $r_2/b= 0.1$ as a function of the 
cutoff in the dimensionless combination $\Lambda b$.}
\label{C4_lambda5_b_a_1}
\end{figure}

As an illustration, we take $r_2/b= 0.1$.
The coupling constants for finite scattering length $a_2/b=1$
are plotted in 
Figs. \ref{C0_lambda_b_a_1}, \ref{C2_lambda3_b_a_1}, 
and \ref{C4_lambda5_b_a_1}.
The running of $C_0^{(0)}$ is identical to the one in the previous
subsection, and as advertised is gentler than the one displayed in 
Fig. \ref{C0_lambda_unitarity}.
As seen in Eqs. (\ref{C01asym}) and (\ref{C21asym}), the range 
changes the running of $C_0^{(1)}$ and $C_2^{(1)}$
dramatically with respect to the zero-range case, say 
Figs. \ref{C0_lambda_unitarity} and 
\ref{C2_lambda3_unitarity}. 
The new runnings approach the limits given
in Eqs. (\ref{C01asym}) and (\ref{C21asym}).
At $\Lambda b\simeq 38$ we find, for example, 
$8\pi C_2^{(1)}/ \mu r_2 C_0^{(0)}\simeq 0.7$.
The slow convergence is a consequence that at this cutoff
$r_2\Lambda$ is only $\simeq 3.8$, and
the ${\cal O}(1/r_2\Lambda)$ corrections are still not so small.
For the other coupling constants, $C_0^{(2)}$, $C_2^{(2)}$, and $C_4^{(2)}$, 
the running
is also very different from the zero-range case. 
In agreement with our error estimates, these parameters
are nearly flat when normalized with $1/r_2^2$.

\begin{figure}[tb]
\begin{center}
\includegraphics*[scale=0.82]{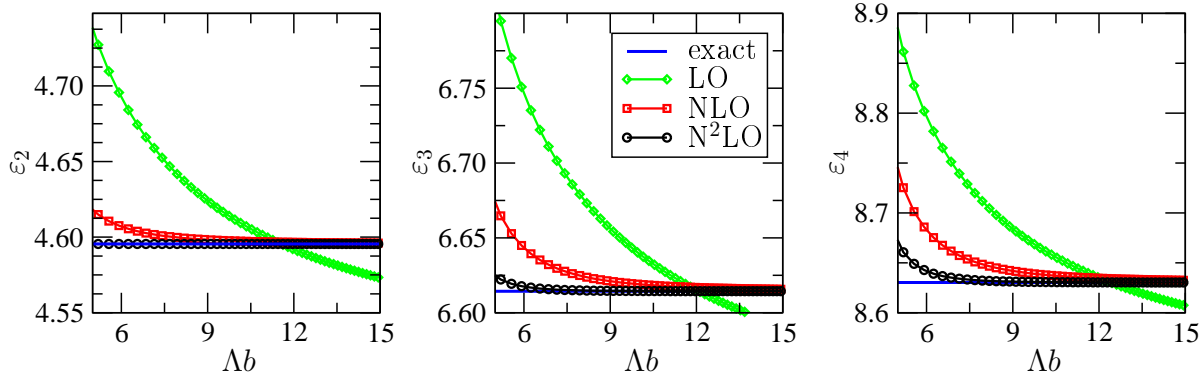}
\end{center}
\caption{Same as in Fig. \ref{states_unitarity}, but for $r_2/b=0.1$.
}
\label{states_unitarity_point_one}
\end{figure}

\begin{figure}[tb]
\begin{center}
\includegraphics*[scale=0.82]{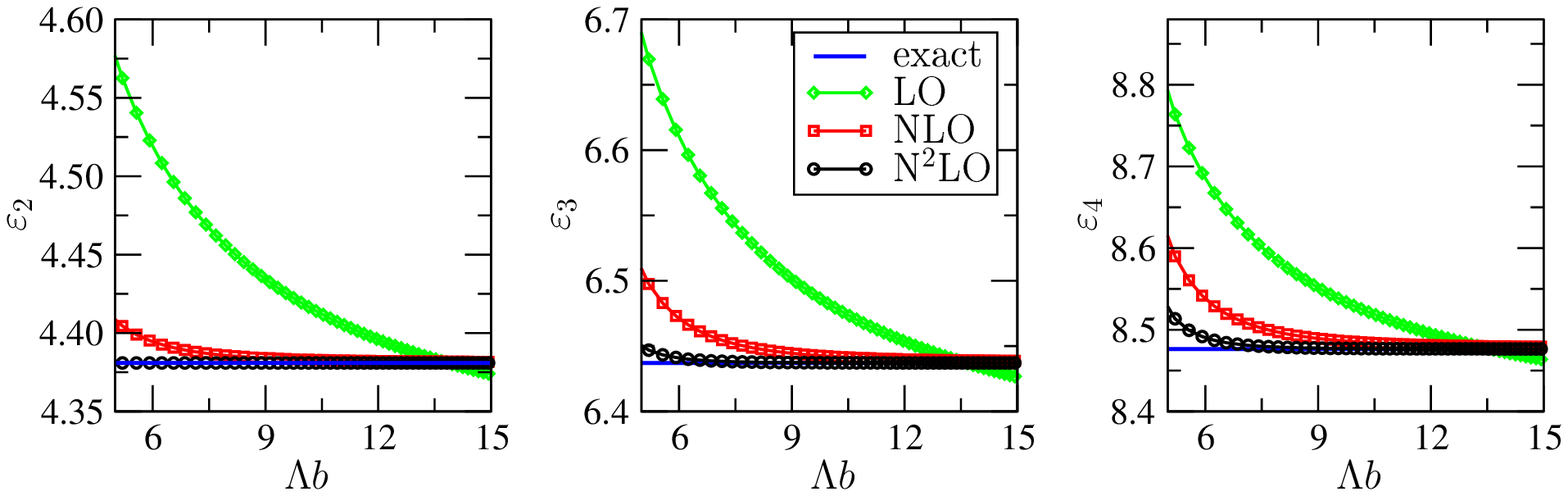}
\end{center}
\caption{Same as in Fig. \ref{states_unitarity}, 
but for $b/a_2=1$ and $r_2/b=0.1$.}
\label{states_b_over_a_1_point_one}
\end{figure}

Qualitatively, the only change in energies with respect to previous
subsections is the finite jump from LO
to NLO due to the effective range.
Energies for a few excited states 
in the case of  an infinite scattering length, $b/a_2=0$,
are plotted in Fig. \ref{states_unitarity_point_one},
and for a finite scattering length, $b/a_2=1$,
in Fig. \ref{states_b_over_a_1_point_one}.
As evident, LO misses the correct asymptotic behavior by a little bit
because it lacks the effective range. 
The NLO results converge, as $n_{max}\to \infty$, to the values 
given by Eq. (\ref{transcend}) with the range, as they should.
However, the inclusion of N$^2$LO corrections speeds up convergence 
considerably:
for a fixed value of the cutoff $n_{max}$ the results at N$^2$LO are 
much closer to the exact value.
It is straightforward, for example, to generalize 
Eqs. (\ref{deutBE}) and (\ref{deltaeigv_loIII}) to non-zero range.

\section{Conclusions and Outlook}
\label{sect4}
We have extended the work initiated in Ref. \cite{trap0} on the application of 
effective field theory to the trapped two-particle system to include
subleading orders.
In EFT, the short-range interparticle potential is replaced by a 
delta function
and its derivatives, as we presented in the main text, or equivalently
by an energy-dependent delta function, as discussed in App. \ref{AppD}.
The singular nature of the interaction requires regularization and
renormalization, which is naturally accomplished by the use of a finite
model space, as required in actual calculations.
Once the parameters of the interaction are fixed in each model
space using a finite number of levels, other levels can be
calculated.

The leading-order interaction is solved for exactly, 
whereas higher contributions are treated in perturbation theory. 
We have 
considered the corrections to the potential up to 
next-to-next-to-leading-order. 
We have shown explicitly that this method can systematically account
for the physics of the effective-range expansion,
while treating the subleading potential exactly gives worse results.

In the limit of a large model space, the leading-order theory
reproduces the pseudopotential result of Ref. \cite{HT},
while in subleading orders our approach gives range and higher ERE corrections
\cite{models,DFT_short,mehen} to the pseudopotential result.
Therefore we have provided an alternative derivation of these results.
In turn, these asymptotic results allow us to incorporate
the physics of scattering 
into the trapped system: by fitting the coupling
constants to the asymptotic values of some levels,
we can calculate other energies without necessarily resorting to fitting
measured bound-state energies.  
This is important in the nuclear case where the scattering parameters are 
known.

We have studied numerically 
the convergence to large model spaces in some detail.
We have presented results at unitarity and for finite values of the scattering 
length $a_2$ as well as for finite values of the effective range $r_2$. 
In all cases, 
we have observed 
convergence to the asymptotic values.
Moreover, for a fixed value of the cutoff, the difference between the exact 
value and the value obtained in a finite model space is mitigated as more 
corrections to the potential are taken into account.
Truncation errors can thus be reduced not only by increasing the 
size of the model
space but also by increasing the order of the calculation.

Thus, we have a method to calculate the energies of few-body systems
that is independent of the details of the short-range potential
and can be improved systematically. 
Although in the two-body system our method is simply an implementation
of the effective-range expansion, it is well-suited for extension
to larger systems where numerical calculations are required.
In these cases, limited computational power restricts
the size of accessible model spaces. We can fight this limitation by 
increasing the order of the calculation, thus accelerating convergence.
Results for systems with more particles will be presented in a
future publication \cite{usnext}.

\section*{Acknowledgments}
We thank Thomas Papenbrock for pointing out some useful references
and Mike Birse for interesting discussions.
The work reported here benefited from hospitality extended to its
authors by the National Institute for Nuclear Theory 
at the University of Washington during the Program on Effective Field Theories
and the Many-Body Problem (INT-09-01),
and to UvK by the Kernfysisch Versneller Instituut at the
Rijksuniversiteit Groningen.
This research was supported in part by 
NSF grants PHY-0555396 and PHY-0854912 (BRB, JR), and
by US DOE grants 
DE-FC02-07ER41457 (IS)
and 
DE-FG02-04ER41338 (JR, UvK).

\appendix

\section{Potential}
\label{AppD}

Analogously to the multipole expansion in electrodynamics,
the short-range two-body potential can be expanded in a power series
in momenta \cite{pbc98,aleph},
\begin{eqnarray}
V(\vec{p}\,', \vec{p}) &=& C_0 + C_2 \left(\vec{p}\,'^2 + \vec{p}\,^2\right)
+ C_4 \left(\vec{p}\,'^2 + \vec{p}\,^2\right)^2 
+ \tilde{C}_{4} \left(\vec{p}\,'^2 - \vec{p}\,^2\right)^2 
\nonumber\\
&&+ C_2' \vec{p}\,'\cdot \vec{p} 
+ C_4' \vec{p}\,'\cdot \vec{p} \left(\vec{p}\,'^2 + \vec{p}\,^2\right)
+ C_4'' (\vec{p}\,'\cdot \vec{p})^2
+\ldots
\label{Taylor}
\end{eqnarray}
where $\vec{p}$ ($\vec{p}\,'$) is the initial (final) relative momentum,
the $C_i^x$  are constants,
and ``$\ldots$'' denote terms with six or more powers
of momenta.
Here the $C_0$, $C_2$, and $C_4$ terms contribute to the 
$S$ wave,
$C_2'$ and $C_4'$ to the $P$ wave, and $C_4''$ to the $D$ wave.
The contribution from $\tilde{C}_{4}$ to the on-shell two-body system
vanishes, and thus it is only relevant in larger systems, where it cannot
be separated from few-body forces.
If we were to include this term and 
set a similar linear system to Eq. (\ref{3eqs})
(based on the fit of the first four levels inside the HO),  
this system would have no solution:
its zero determinant would indicate that
the corrections $C_4$ and $\tilde{C}_4$ are not independent.

In the situation of interest here, where the two-body $S$-wave
scattering length $a_2$ is large,
the $S$-wave constants are enhanced over the others
by powers of $a_2$, and as a result
the $C_0$ term is LO,
the $C_2$ term is NLO, 
the $C_4$ term is N$^2$LO,
while the others appear only at higher orders \cite{aleph}.
In this paper we limit ourselves to N$^2$LO, where only the $S$ wave
is present. 
Taking the Fourier-transform with respect to both $\vec{p}$ and $\vec{p}\,'$,
the potential 
is found to be given in coordinate space by Eq. (\ref{Taylorcoord}).
It is non-local in the sense of involving derivatives.

A completely equivalent formulation of the EFT in the two-body 
system is achieved with an energy-dependent potential \cite{aleph}.
It can be implemented through an auxiliary ``dimeron'' field \cite{dibaryon}.
In LO, the dimeron is characterized by a mass $\Delta$
and a coupling constant $g$ to a two-particle $S$-wave state.
In the case of interest here, $|a_2|\gg R$, the kinetic energy
of the dimeron is an NLO effect \cite{aleph}.
A subtlety is that the bare dimeron can be a ghost:
the sign $\sigma$ of the kinetic term can be positive or negative, 
depending on the sign of the effective range.
Denoting by $E$ is the total energy in the center-of-mass frame,
the two-body potential is simply
\begin{eqnarray}
V(E) &=& -\frac{g^2}{\Delta} - \sigma \frac{g^2}{\Delta^2} E 
+\ldots
\label{EdepV}
\end{eqnarray}
in momentum space, and
\begin{eqnarray}
V(\vec{r}\,', \vec{r})
&=&-\frac{g^2}{\Delta} \left(1+\sigma \frac{E}{\Delta} +\ldots\right)
\delta(\vec{r}\,') \delta(\vec{r})
\label{TaylorcoordEdep}
\end{eqnarray}
in coordinate space. It is local, but energy dependent.

This potential was considered in Ref. \cite{mehen}.
An alternative treatment 
follows the same steps as that
of the
momentum-dependent potential (\ref{Taylorcoord}) in the main text, 
with the substitutions
\begin{eqnarray}
C_0^{(0)}&\to & -\frac{g^{(0)2}}{\Delta^{(0)}},\label{sub1}\\
C_0^{(1)}&\to & -\frac{g^{(1)2}}{\Delta^{(1)}}
\left(1+\sigma \frac{E}{\Delta^{(1)}}\right),\label{sub2}\\
C_2^{(1)}&\to &  0\label{sub3}\\
&\ldots & \nonumber
\end{eqnarray}
In particular, the LO is identical to that presented in the main text,
with Eq. (\ref{sub1}). 
Note that since only the ratio $g^{(0)2}/\Delta^{(0)}$ enters,
the separation between $g^{(0)}$ and $\Delta^{(0)}$ is arbitrary.
At NLO, one finds equations that are 
somewhat simpler than 
Eq. (\ref{eq:corrNLO}) and the ones that follow.
With Eqs. (\ref{sub2}) and (\ref{sub3}) one can find 
the renormalization of $g^{(1)}$ and $\Delta^{(1)}$ separately.
The appearance of energy instead of momentum leads to a
softening of the UV behavior. 
This results in a wavefunction at NLO of exactly the form 
(\ref{wf2_NLO}) and (\ref{A1}),
but with Eq. (\ref{sub3}).
In the infinite-cutoff limit the most singular term
in Eq. (\ref{wf_NLO_inf}) is absent, and 
Eq. (\ref{expansion_NLO}) follows directly.
Any observable in the two-body system is identical for this
potential and the momentum-dependent potential discussed in the text.

\section{HO notation and definitions}
\label{appA}

The HO basis functions with the length parameter
\begin{equation}
 b=\frac{1}{\sqrt{\mu\omega}}
\end{equation}
are the solutions of the three-dimensional 
Schr\"odinger equation 
\begin{equation}
\frac{1}{2} \left(-b^2\nabla^2+\frac{r^2}{b^2}\right)\phi_{nlm}(\vec r)
=\epsilon_{nl} \phi_{nlm}(\vec r)
\end{equation}
with energy (in units of $\omega$)
\begin{equation}
\epsilon_{nl}=\frac{E_{nl}}{\omega}= 2n+l+\frac{3}{2}.
\end{equation}
They are given by
\begin{equation}
\langle \vec{r}\, |nlm\rangle 
= \phi_{nlm}(\vec r)=R_{nl}(r)Y_{lm}(\hat r),
\end{equation}
where $Y_{lm}(\hat r)$ are the usual spherical harmonics, 
and the radial parts $R_{nl}(r)$ can be shown to have the form
\begin{eqnarray}
R_{nl}(r)&=&
\left(\frac{2}{b^3}\frac{1}{\Gamma(l+3/2)}\right)^{1/2}
\left[L_n^{(l+1/2)}\left(0\right)\right]^{-1/2}
\left(\frac{r}{b}\right)^l
\exp\left(-r^2/2b^2\right)
L_n^{(l+1/2)}\left(r^2/b^2\right).
\label{defHOwf}
\end{eqnarray}
The $L_n^{(\alpha)}(x)$'s are the generalized Laguerre polynomials,
which can be written as \cite{abramowitz}
\begin{equation}
L_n^{(\alpha)}(x) =
\frac{\Gamma(n+\alpha+1)}{n!\; \Gamma(\alpha+1)} 
M\left(-n, \alpha + 1, x \right)
\label{laguerregen}
\end{equation}
in terms of the confluent hypergeometric function $M$.

For contact interactions, it is useful to know the value of the 
radial wavefunction at the origin. 
Because of the $r^l$ factor, only $l=0$ contributes, as expected. 
The $S$ wavefunction of energy $E_n=(2n+3/2)\omega$ is,
omitting the $l=m=0$ labels,
\begin{eqnarray}
\langle \vec{r}\, |n\rangle  = \phi_{n}(r) =
\pi^{-3/4} b^{-3/2}
\left[L_n^{(1/2)}\left(0\right)\right]^{-1/2}
\exp\left(-r^2/2b^2\right)
L_n^{(1/2)}\left(r^2/b^2\right).
\label{HOSwfprime}
\end{eqnarray}
In particular,
\begin{equation}
 \phi_{n}(0)=\pi^{-3/4} b^{-3/2}
\left[L_n^{(1/2)}\left(0\right)\right]^{1/2},
\end{equation}
where
\begin{equation}
L_n^{(\alpha)}(0) 
=\frac{\Gamma(n+\alpha+1)}{n!\; \Gamma(\alpha+1)}
=\prod_{k=1}^n \left(1+\frac{\alpha}{k}\right)
=\left(1+\frac{\alpha}{n}\right) L_{n-1}^{(\alpha)}(0).
\label{laguerre}
\end{equation}
Use of Stirling's formula yields
\begin{equation}
\frac{\Gamma(z+\alpha)}{\Gamma(z)}= z^\alpha 
    \left\{1+\frac{\alpha(\alpha -1)}{2z} +O(z^{-2})\right\}
 \label{asyGr}
\label{stirling}
\end{equation}
for large $z$, which leads to
\begin{equation}
L_n^{(\alpha)}(0) = \frac{n^\alpha}{\Gamma(\alpha+1)}  
                    \left\{1+\frac{\alpha(\alpha+1)}{2n}+O(n^{-2})\right\}
\label{Llargen}
\end{equation}
for large $n$.

The generalized Laguerre polynomials satisfy \cite{HT}
\begin{equation}
\sum_{n=0}^{\infty} \frac{L_n^{(1/2)}(x)}{n+a}
=\Gamma{(a)} U\left(a,3/2, x\right),
\label{sumL}
\end{equation}
in terms of the confluent hypergeometric function \cite{abramowitz}
\begin{equation}
U\left(a,3/2,x\right)=\sqrt{\frac{\pi}{x}} \left[
    \frac{M\left(a-1/2,1/2,x\right)}{\Gamma(a)\Gamma(1/2)}
   -\frac{\sqrt{x} M\left(a,3/2,x\right)}{\Gamma(a-1/2)\Gamma(3/2)}\right].
\end{equation}
For small $x$,
\begin{equation}
\Gamma{(a)} U\left(a,3/2, x\right)= \sqrt{\frac{\pi}{x}} 
\left[1-2\frac{\Gamma(a)}{\Gamma(a-1/2)}\sqrt{x}+O(x)\right].
\label{smallx}
\end{equation}
Also useful are sums involving the generalized Laguerre polynomials at the 
origin \cite{wolframalpha}:
\begin{equation}
\sum_{n=0}^{m} L_n^{(1/2)}(0)
=\frac{4}{3\sqrt{\pi}} \frac{\Gamma(m+5/2)}{\Gamma(m+1)},
\label{sumL03}
\end{equation}
\begin{equation}
\sum_{n=0}^{m} \frac{L_n^{(1/2)}(0)}{n+1/2}
=\frac{4}{\sqrt{\pi}} \frac{\Gamma(m+3/2)}{\Gamma(m+1)},
\label{sumL01}
\end{equation}
\begin{eqnarray}
&&\sum_{n=0}^{m} \frac{L_n^{(1/2)}(0)}{(n+1/2)(n+a)}
=2\sqrt{\pi}\left[
\frac{\Gamma(a)}{\Gamma(a+1/2)}
\right.
\nonumber\\
&&\left.-\frac{\Gamma(m+3/2)}{\Gamma(m+2)}
\, \frac{_3F_2\left(1, m+3/2, a+m+1; m+2, a+m+2; 1 \right)}
        {\pi(a+m+1)}
\right],
\label{sumL02}
\end{eqnarray}
and
\begin{eqnarray}
&&\sum_{n=0}^{m} \frac{L_n^{(1/2)}(0)}{(n+1/2)(n+a)^2}
=-2\sqrt{\pi}\left\{
\frac{\Gamma(a)}{\Gamma(a+1/2)}
\left[\psi^{(0)}(a)-\psi^{(0)}(a+1/2)\right]
\right.
\nonumber\\
&&\left.
+\frac{\Gamma(m+3/2)}{\Gamma(m+2)}
\,
\frac{_4F_3\left(1, m+3/2,a+m+1,a+m+1;
      m+2,a+m+2,a+m+2;1\right)}{\pi(a+m+1)^2} 
\right\},
\nonumber\\
\label{sumL04}
\end{eqnarray}
where $_3F_2$ and $_4F_3$ are generalized hypergeometric functions
and $\psi^{(0)}=\Gamma'/\Gamma$ is the digamma function \cite{abramowitz}.

\section{Wavefunction at NLO} 
\label{wf_NLO}
The correction at NLO is taken into account as a perturbation to the LO 
potential.
We write the corresponding quantum state of the two-fermion system as 
$|\psi\rangle=|\psi^{(0)}\rangle+|\psi^{(1)}\rangle$
where 
$|\psi^{(0)}\rangle$ is the solution at LO and 
$|\psi^{(1)}\rangle$ the correction at NLO. 
Both of these vectors are expanded in a HO basis,
as in Eq. (\ref{expansionS}).

Corrections at NLO are obtained by 
solving the 
Schr\"odinger equation at first order, Eq. (\ref{sch1}).
Let us introduce the projection $R_n$ of the left and right sides of 
Eq. (\ref{sch1}),
\begin{eqnarray}
R_n\equiv \langle n |(H^{(0)}-E^{(0)})|\psi^{(1)}\rangle \label{R_n_l} 
=\langle n |(E^{(1)}-V^{(1)})|\psi^{(0)} \rangle \label{R_n_r},
\label{R_m2_app}
\end{eqnarray}
and define $\kappa^{(1)}$ as $\kappa^{(1)}=C_0^{(0)} \psi^{(1)}(0)$.
{}From the lhs of Eq. (\ref{R_n_l}) we can write the coefficient of the wave 
function  at NLO as
\begin{eqnarray}
c_{n}^{(1)}= \frac{-\kappa^{(1)} \phi_n(0)+ R_n} {E_n-E^{(0)}}.
\label {cm1_app}
\end{eqnarray}

Normalizing the wavefunction at LO to unity,
\begin{eqnarray}
\kappa^{(0)-2} =  \sum_{n} \frac{\phi_{n}^{2}(0)}{(E_n-E^{(0)})^{2}}.
\label{normLOh}
\end{eqnarray}
Keeping this normalization at NLO,
$|\psi^{(0)}\rangle$ and $|\psi^{(1)}\rangle$ are orthogonal;
one then obtains, from Eqs. (\ref{cm}) for $c_{n}^{(0)}$
and (\ref{cm1_app}) for $c_{n}^{{(1)}}$, 
\begin{eqnarray}
\langle \psi^{(0)}|\psi^{(1)}\rangle &=& \sum_{n} c_{n}^{(0)} c_{n}^{(1)}
= \frac{\kappa^{(1)}}{\kappa^{(0)}}
-\kappa^{(0)}\sum_{n} \frac{\phi_n(0)R_n}{(E_n-E^{(0)})^2}=0.
\label{ortho}
\end{eqnarray}

On the other hand, the rhs of Eq. (\ref{R_m2_app}) provides
an expression for $R_n$.
Using Eq. (\ref{cm}),
\begin{eqnarray}
[\nabla^2 \psi^{(0)}(r)]_{r=0}= 
-2\mu \sum_{n} c_{n}^{(0)} E_n\phi_{n}(0)
= 2 \mu \left[ -E^{(0)} \psi^{(0)}(0) 
+\kappa^{(0)} \sum_{n} \phi_{n}^{2}(0)\right].
\label {laplace_app}
\end{eqnarray}
Inserting this expression into the rhs of Eq. (\ref{R_m2_app}) and 
eliminating $\psi^{(0)}(0)$,
\begin{eqnarray}
R_n =  \phi_{n}(0) \frac{\kappa^{(0)} }{ C_0^{(0)}}  
\left[ \frac{E^{(1)} C_0^{(0)}} {E^{(0)}-E_n} -C_0^{(1)} 
+2 \mu C_2^{(1)} \left ( E_n+E^{(0)}-C_0^{(0)}\sum_{m}\phi_{m}^{2}(0) 
\right) \right].
\label{R_m2_ter_app}
\end{eqnarray}
{}From Eqs. (\ref{ortho}) and (\ref{R_m2_ter_app}), 
we arrive at an expression for $\kappa^{(1)}$ using Eq. (\ref{C_0_bis}):
\begin {eqnarray}
\kappa^{(1)} &=& \kappa^{(0)3}
\left [E^{(1)} \sum_{n} \frac{\phi_n^2(0)}{(E^{(0)}-E_n)^3} 
-\frac{C_0^{(1)}}{C_0^{(0)}} 
 \sum_{n} \frac{\phi_n^2(0)}{(E^{(0)}-E_n)^2} \right] 
\nonumber \\ 
&&
+2 \mu C_2^{(1)} \left[\left(\frac{\kappa^{(0)}}{C_0^{(0)}}\right)^2  
+\frac{2E_0}{C_0^{(0)}} -\sum_{n} \phi_{n}^{2}(0)\right] .
\label{kappa1_app}
\end{eqnarray}
We can now finally obtain the coefficient $c_n^{(1)}$ by plugging
Eqs. (\ref{R_m2_ter_app}) and (\ref{kappa1_app}) into Eq. (\ref{cm1_app}):
\begin{eqnarray}
c_n^{(1)}&=&
c_n^{(0)}\left[\frac{E^{(1)}}{E_n-E^{(0)}}
+ \kappa^{(0)2} E^{(1)}   
\sum_m \frac{\phi_{m}^{2}(0)}{(E^{(0)}-E_m)^3} 
+2 \mu C_2^{(1)} 
\left(\frac{\kappa^{{(0)}}}{C_0^{(0)}} \right)^2\right]  
\nonumber \\
&&
+ 2\mu \kappa^{(0)} \frac{C_2^{(1)}}{C_0^{(0)}}\phi_n(0).
\end{eqnarray}

The total wavefunction at NLO is then given by
\begin{eqnarray}
\psi(r)&=&\sum_{n} (c^{(0)}_n +c^{(1)}_{n})\phi_n(r) \\
&=& \sum_{n} c^{(0)}_n\left (1+A^{(1)}+\frac{E^{(1)}}{E_n-E^{(0)}} \right ) 
\phi_n(r)
+2\mu \kappa^{(0)}  \frac {C_2^{(1)}}{C_0^{(0)}}
\sum_{n} \phi_n(0)\phi_n(r),
\end{eqnarray}
where $A^{(1)}$  is a first-order term defined as
\begin{eqnarray}
A^{(1)}=\kappa^{(0)2} \left[ E^{(1)}    
\sum_m \frac{\phi_{m}^{2}(0)}{(E^0-E_m)^3}
+2\mu \frac{C_2^{(1)}}{C_0^{(0)2}} \right].
\end{eqnarray}
At first order in $A^{(1)}$ and $E^{(1)}$, the previous expression 
for the wavefunction is equal to
\begin{eqnarray}
\psi(r)
&=&\left (1+A^{(1)}\right )  \kappa^{(0)}
\sum_{n} \frac{\phi_n(0)\phi_n(r)}{E^{(0)}+E^{(1)}-E_n}
+2\mu  \kappa^{(0)} \frac{C_2^{(1)}}{C_0^{(0)}} 
\sum_n \phi_n(0)\phi_n(r).
\end{eqnarray}
Upon insertion of the expression for the radial wavefunction 
of the HO basis we obtain Eqs. (\ref{wf2_NLO}) and (\ref{A1}).

\end{document}